\documentclass[twocolumn,showpacs,preprintnumbers,amsmath,amssymb,prb]{revtex4}

\usepackage{graphicx}
\usepackage{dcolumn}
\usepackage{bm}

\begin{document}

\preprint{Physica C, Special Issue on Superconducting Pnictides}

\title{
Frustrated Magnetic Interactions,  Giant  Magneto-Elastic 
Coupling, and Magnetic Phonons in Iron-Pnictides}


\author{Taner  Yildirim$^{1,2}$}\email{taner@nist.gov}%
\affiliation{%
$^{1}$NIST Center for Neutron Research, National Institute of Standards and
Technology, Gaithersburg, Maryland 20899, USA
\\$^{2}$Department of Materials Science and Engineering, University of
Pennsylvania, Philadelphia, PA 19104, USA}%

\date{\today}

\begin{abstract} 
We present a detailed first principles study of Fe-pnictides with particular emphasis on
competing magnetic interactions, structural phase transition,  giant magneto-elastic
coupling and its effect on phonons.  The exchange interactions $J_{i,j}(R)$ are calculated up to
$\approx 12 \;$\AA $\;$ from two different approaches based on direct spin-flip and infinitesimal spin-rotation. 
We find that $J_{i,j}(R)$ has an oscillatory
character with an envelop decaying as $1/R^3$ along the stripe-direction while 
it is very short range along the diagonal direction and antiferromagnetic.
A brief discussion of the neutron scattering determination of these exchange constants from a 
single crystal sample with orthorhombic twinning is given.
 The lattice parameter dependence of the exchange 
constants, $dJ_{i,j}/da$ are calculated for a simple   spin-Peierls like  
model to explain the fine details of 
the tetragonal-orthorhombic phase transition.
We then discuss  giant  magneto-elastic effects in these systems.
We show that when the Fe-spin  is turned off  the optimized c-values are 
shorter than experimetnal values by 1.4 \AA $\;$
for CaFe$_2$As$_2$, by 0.4 \AA $\;$ for BaFe$_2$As$_2$, and by 0.13 \AA $\;$ for LaOFeAs.
We explain this strange behavior by unraveling surprisingly strong
interactions between arsenic ions, the strength of which  is  
controlled by the Fe-spin state through Fe-As hybridization. 
Reducing the Fe-magnetic moment, weakens the Fe-As bonding, and in turn, 
increases As-As interactions, causing a giant reduction in the c-axis.
These findings also explain why the Fe-moment is so tightly coupled
to the As-z position. Finally, we show that Fe-spin is also required to obtain the right phonon
energies, in particular As  c-polarized and Fe-Fe in-plane modes that have been recently observed by inelastic 
x-ray and neutron scattering  but cannot be explained based on 
non-magnetic phonon calculations.
Since treating iron as magnetic ion always gives much better results than non-magnetic ones and
since there is no large c-axis reduction during the normal to superconducting 
phase transition, the iron magnetic moment should be present in Fe-pnictides at all times.
We discuss the implications of our results on the mechanism of  superconductivity 
in these  fascinating Fe-pnictide systems.
\end{abstract}

\maketitle

\section{Introduction}

The recent discovery of superconductivity at 
T$_c$'s up to 55~K in iron-pnictide systems\cite{kamihara,sm_43k,ce_41k,pr_52k}
has sparked enormous interest in this class of materials. So far four types of materials
have been discovered. The first one is the rare-earth  pnictide oxide layered systems, 
REOFeAs which is denoted as "1111"\cite{kamihara,sm_43k,ce_41k,pr_52k,afe2as2_refeas}. 
The second class is the so called "122"
systems with the chemical formula  
MFe$_2$As$_2$ (M=Ca,Sr, etc)\cite{bafe2as2,srfe2as2,cafe2as2,srfe2as2_sc,afe2as2_refeas}. 
The third system with $T_c=18$ K is MFeAs (M=Li and Na), which is similar to 
REOFeAs but instead of REO-layers, we have
now small alkali metals such as Li\cite{lifeas}. The last one is 
the binary Fe(Se,Te) systems which have been shown to superconduct up to 12 K under
pressure\cite{fese}.

\begin{figure}
\includegraphics[width=8cm]{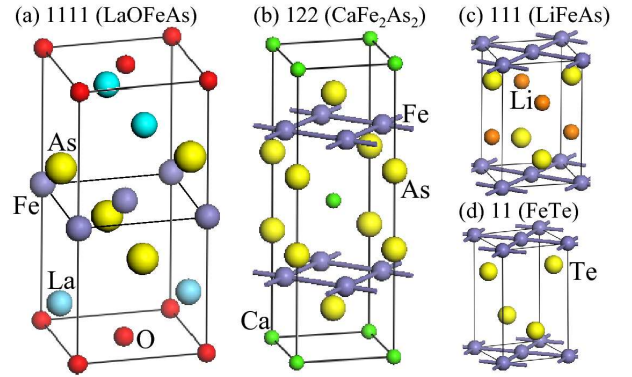}
\caption{
(color online)
 The crystal structures (with origin choice 1) of four types of 
 Fe-pnictide systems that have been discovered so far.
 }
\label{fig1}
\end{figure}

The crystal structures of these four systems are shown
in  Fig.~1. The common feature in these Fe-pnictide superconductors is the
presence of FeAs plane (or Fe(Te,Se) in the case of 11 systems),  which is shown in Fig.~2.
Basically Fe atoms form a regular square lattice just like the CuO$_2$ plane in cuprates. 
However the important difference is the location of the arsenic ions which are not located between
two Fe ions but rather above/below the center of Fe-square. This arrangement of arsenic ions  has 
several important consequences in the electronic and magnetic properties of these systems. 
Since As is not directly between two Fe ions, the Fe-Fe 
distance is not large and direct Fe-Fe overlap plays an important role 
in the band formation near the Fermi level. Then, the delicate interplay between Fe-Fe, Fe-As,
and even As-As interactions (which is very important in 122 systems such as Ca122) 
result interesting electronic and magnetic properties that are super-sensitive to the 
As-z position and the c-lattice parameter of the
Fe-pnictide system.
In this paper, we will focus on the structural, dynamical and magnetic properties of  1111 and 122 systems only. 
For a recent review of 
electronic  and superconducting properties of  Fe-based superconductors,  we refer the reader to 
David Singh's \cite{david_review}  and Igor Mazin's \cite{igor_review2} articles
in this issue and  references therein.


 A common phase diagram for iron-pnictides has emerged\cite{jeff_review} in which the 
 stoichiometric parent compound shows a structural anomaly around 150-200 K, 
  below which spin-density-wave (SDW) antiferromagnetic 
  ordering\cite{cruz,laofeas_local,srfe2as2,cafe2as2,srfe2as2_sc} 
 appears, which is due to nesting Fermi surfaces\cite{ishibashi,dong,mazin,singh}.  
 The SDW ordering is further
 stabilized against the normal checkerboard antiferromagnetic ordering (denoted as AF1) due to strong
 antiferromagnetic interactions along the Fe-square diagonal\cite{yildirim_prl1}. 
   Superconductivity in these systems only  occurs when the SDW ordering
  and the structural distortion are suppressed, which can be achieved in 
  a number of ways such as fluorine doping on the oxygen 
  site\cite{kamihara,sm_43k,ce_41k,pr_52k}, or
 hole doping (La$_{1-x}$Sr$_x$)\cite{sr_doped,srkxfe2as2,srcsxfe2as2,canaxfe2as2}  
 or by applying  external pressure\cite{pres_cafe2as2_sc,sr_ba_fe2as2_pres,ph_tr_cafe2as2}. 
 
 The structural distortion which is common to all parent compounds can be characterized by either 
 primitive monoclinic space group P2/c (P112/n) or the conventional orthorhombic cell with space group
 Cmma\cite{cmma}. The relation of these two representations is indicated in the 
  bottom panel of Fig.~2. We note that when the system is distorted (i.e. $\gamma \neq 90.0 $)
  one of the Fe-pairs gets closer and the other Fe-Fe distance gets longer, yielding an
  orthorhombic lattice (i..e $a_o \neq b_o$).  Below we will successfully explain how this structural phase
  transition is tightly coupled to the magnetic SDW ordering. Finally
  we note that the reported space groups in the SDW state\cite{cruz,cmma} (i.e. Cmma or P2/c) 
  are actually  the space groups  of the system without the Fe-magnetic moment. 
  Hence technically the Cmma is not the right space group for the SDW magnetic system. 
  If we ignore the antiferromagnetic stacking of the FeAs planes, the actual space group is 
  Pbnm which is primitive as expected. This will be important
  in the discussion of magnetic phonon calculations in Sec. VI where we can not use the 
  space group Cmma but has to use Pbmb for 1111 systems. The situation is very similar for 
  the 122, 111, and 11 systems as well. For example, the Fmmm space group of 111-systems is actually 
  reduced to Bbmb (spg. number=66, origin 1) when the iron-spins are considered.

  Clearly, the understanding of electronic, magnetic, and structural properties of the parent FeAs 
 compound
is the key to determining the underlying mechanism that makes these materials superconduct upon 
electron/hole doping. 
 In this paper we present  a detailed  first-principles study of Fe-pnictides 
 with main focus on the competing magnetic spin-interactions, structural phase transition,
the  giant magneto-elastic coupling and the phonons. 
 Our main objective is to demonstrate that Fe-spin is the key in understanding many properties of these
 systems, including lattice parameters, atomic positions, and the  phonon spectrum.
 When the Fe-spin is ignored and non-magnetic calculations are done, the results do not agree
 with most of the experimental data. This observation could be the key in identifying the mechanism of 
 superconductivity in these systems.

\begin{figure}
\includegraphics[width=6cm]{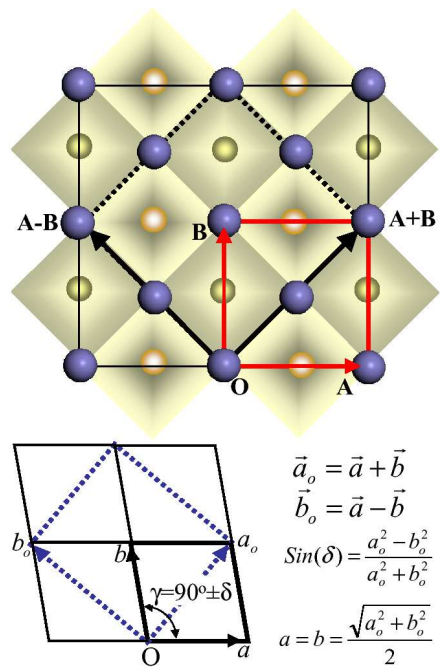}
\caption{
(color online)
 Top: A view along c-axis of the FeAs-plane and the relations between  the primitive and 
$\sqrt{2}\times\sqrt{2}$ supercell used in our calculations. The dark and light shaded areas
indicate the As atoms below and above the Fe-square lattice, respectively.
Bottom: Relation between conventional (Cmma) and primitive (P2/c) cells of 
the orthorhombic  structure.
 }
\label{fig2}
\end{figure}

 This paper is organized as follows.  In the next section, we discuss the energetics of possible
 spin configurations in Fe-pnictides within a unified model from all-electron
 fix-spin moment calculations. 
 We will show that the SDW magnetic ordering is the only stable ground state
 for Fe-pnictide.  In Sec. III, we will calculate
 the exchange interactions $J_{i,j}(R)$ up to
$\approx 12 \AA\;$ using two different approaches based on direct spin-flip and infinitesimal spin-rotation. 
We find that $J_{i,j}(R)$ has an oscillatory
character with an envelop decaying as $1/R^3$ along the stripe-directions. 
On the other hand,  it is short range along the diagonal direction and antiferromagnetic,
suggesting it is superexchange type and an important contributer
towards the stabilization of SDW ordering. 
A brief discussion of the experimental determination of these exchange constants from an
orthorhombic-twin crystal  is also given in this section.
In Sec. IV, we will discuss
the tetragonal-orthorhombic lattice distortion.
We will calculate the lattice parameter dependence of the exchange 
constants, $dJ_{i,j}/da$, and then use it in a simple 
 spin-Peierls like  
model to explain the fine details of 
the tetragonal-orthorhombic phase transition that is driven by the SDW ordering.
In Sec. V, we will  discuss the giant  magneto-elastic effects in these systems where iron-spin controls the
strength of Fe-As and As-As hybridization which results huge dependence of the magnetic and structural properties on the As-z and
c-axis of the lattice.  Finally, in Sec. VI,  we show that Fe-spin is also required to obtain the right phonon
energies, in particular As  c-polarized and Fe-Fe in-plane modes that have been recently observed by inelastic 
x-ray and neutron scattering measurements but could not been explained based on 
non-magnetic phonon calculations. Our conclusions will be given in Sec. VII.

 \section{Spin Density Wave (SDW) Ordering}

 The early  theoretical studies identified several candidate ground states for
Fe-pnictides such as a nonmagnetic metal near a ferromagnetic or 
antiferromagnetic instability\cite{singh,xu,haule} and a 
simple antiferromagnetic semi-metal\cite{cao,ma}.
The later calculations suggested that Fe-pnictide has
an antiferromagnetic spin-density-wave (SDW) ground state\cite{ishibashi} that is stabilized by
Fermi-surface nesting\cite{dong,mazin} as well as by strong antiferromagnetic spin interactions along the 
Fe-square diagonal\cite{yildirim_prl1,picket}. 
Recently several nice review articles have been published\cite{picket_review,igor_review1}
about the Fermi surface nesting, band structure properties, and the delicate interplay between the structural and electronic
properties\cite{picket_review} in these fascinating systems. Hence,   here we will not discuss the 
Fermi-surface nesting and band structure properties 
but rather focus on the energetics of different spin-configurations in order to determine the nature 
of magnetic interactions present in these systems.

\begin{figure}
\includegraphics[width=8cm]{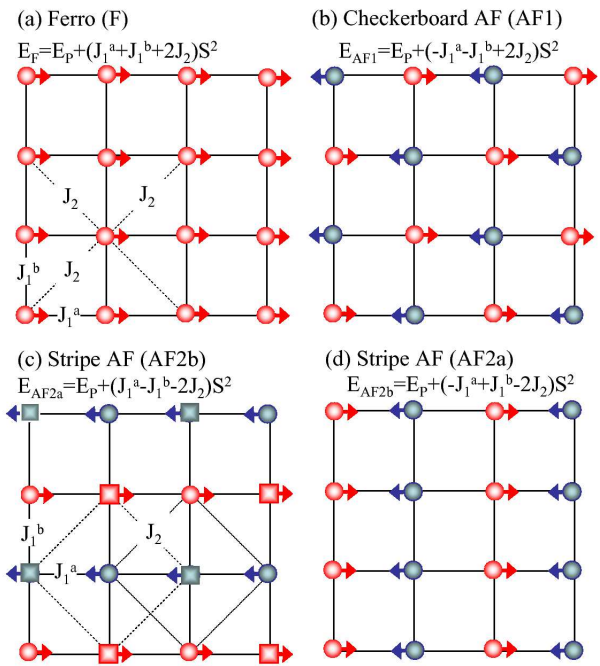}
\caption{
(color online)
 Four possible magnetic configurations for the Fe-square in Fe-pnictide and the corresponding
 energy expressions  in terms of a simple $J_1^a$-$J_1^b$-$J_2$ model.
 Two antiferromagnetic configurations are considered in this study. Top-right  panel (b) shows the AF1 
configuration where nearest neighbor spins are always aligned anti-parallel. 
Two bottom panels (c-d) show the AF2 configuration where the next nearest
neighbor spins (i.e., $J_2$) are always aligned anti-parallel. 
Note that this is the same stripe-phase predicted from
Fermi-surface nesting\cite{dong,mazin} and it is frustrated.
 }
\label{fig3}
\end{figure}

 In order to demonstrate the SDW ground state of Fe-pnictide systems,  one needs to consider
 $\sqrt{2}\times\sqrt{2}$ supercell of the tetragonal cell shown in Fig.~2 with 
 four different magnetic spin configurations.   These are non-magnetic 
 (NM, i.e. no spin polarization), ferromagnetic (F)  and the 
 two different antiferromagnetic spin
 configurations shown in Fig.~3. The first one of the antiferromagnetic configurations is AF1 
 where the nearest neighbor spins are  anti-parallel to each other. 
  The second antiferromagnetic configuration, AF2, is shown in Fig.~3c and 3d.
 In AF2 the Fe spins along the square diagonal are aligned antiferromagnetically. This is the stripe-phase
 which was first  predicted from Fermi-surface nesting\cite{dong,mazin}. The AF2 spin
 configuration can be considered as two interpenerating simple square AF sublattices (circle and square sublattices 
 in Fig.~3c).  From the classical Heisenberg energies of AF1 and AF2, one sees that the
 AF2 spin configuration is stabilized when  $J_2 > J_1/2$. 
 We note that in AF2 spin configuration, due to large antiferromagnetic $J_2$ interactions, the spins along the
 diagonal direction are aligned antiferromagnetically, forcing spins to be parallel and antiparallel along the
 $a-$ and $b-$directions. Hence the $J_1$ interaction can not be fully satisfied and therefore the system is 
 called frustrated. 
 In frustrated magnetic systems, it is known that the
 frustration is usually removed by either a structural distortion or by other effective spin-spin interactions
 that originate from thermal and quantum fluctuations of the spins\cite{yildirim_bct,shlee}. 
 As we shall see below, in Fe-pnictide the Heisenberg picture is only an approximate model. The exchange interactions 
 depend on the spin-configurations considered and  therefore $J_1^b$ could be 
even ferromagnetic in AF2, removing the frustration totally. Finally it has been recently shown that the total
 energy of the AF2 spin configuration increases as $\sin(\theta)^2$ when the two interpenetrating AFM sublattices
 are rotated by an angle $\theta$ with respect to each other\cite{yaresko}. Hence  the infinite degeneracy 
 of the classical Heisenberg model of AF2 structure has been already removed by 
 either  non-Heisenberg like interactions
 and/or long range interactions that are present in Fe-pnictides.

\begin{figure}
\includegraphics[width=4cm]{yildirim_fig3a.eps}
$\;\;$
\includegraphics[width=4cm]{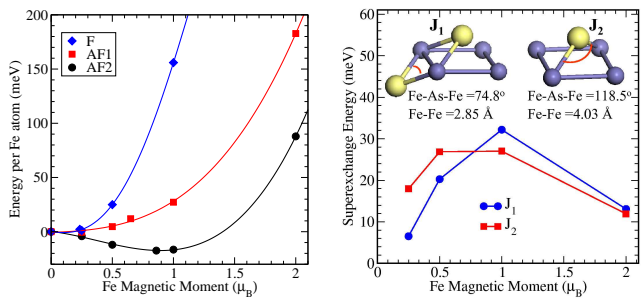}
\caption{
(color online)
(a) The total energy per Fe atom versus magnetic moment for F, AF1 and AF2 spin-configuration,
indicating AF2 is the only ground state of the system. (b) The magnetic interactions for nn and nnn
Fe ions obtained from the energies of F, AF, and AF2 configurations.  
 }
\label{fig4}
\end{figure}

In order to determine which spin configurations among NM, F, AF1, and AF2, is the ground state, 
we have carried out  total energy calculations for each case using experimental structure. 
The calculations were done using the full-potential linearized augmented plane-wave (FP-LAPW) method,
within local density approximation (LDA) using Perdew-Wang/Ceperlye-Alder exchange-correlation\cite{wien2k,exciting}. 
We also used the ultrasoft pseudo potential planewave (PW)
 method\cite{pwscf} for cross checking  of our results and for phonon calculations. 
Since in spin-polarized calculations  it is
very easy to get a local minimum, we followed a different strategy. In our calculations we fixed the
magnetic moment per Fe ion and then  scanned the total energy as a function of Fe-magnetic moment. 
Our results are summarized in Fig.~4. The zero of energy is taken as the
M=0 case (i.e., NM calculation). From Fig.~4, it is clear that LaOFeAs has only one magnetic ground
state which is AF2. The Ferro spin-configuration always results the highest energy regardless  the
Fe-magnetic moment. Similarly AF1 ordering always yields energies higher than the NM case.
For the AF2 ordering, we see that
the energy minimum occurs near the fixed moment calculation with M=1. Repeating calculations where
magnetization is not fixed, we obtained the optimum magnetic moment  as M=0.87 $\mu_B$ per Fe. 
As we discuss below in detail, the Fe magnetic moment is further reduced almost by a half when the structure is
allowed to distort to due to AF2 stripe ordering. 

One confusion with the DFT studies of Fe-pnictide is the calculated  Fe-magnetic moment. The most of the
calculations, in particular those based on pseudopotentials, give too large moment 
around 2.0 $\mu_B$ compared to experimental values of $0.3-0.8$ $\mu_B$. 
In order to explain the small experimental
moment, several theories based on quantum or thermal fluctuations have been proposed 
since the SDW system is magnetically 
frustrated\cite{abraham}. On the other hand, there are studies suggesting that the
 small moment is due to electronic effects,
local chemistry of Fe and its interaction with the As ions.
It has been shown that a small displacement of As-z position and/or structural distortion can 
easily change the Fe-moment 
from 2.0 $\mu_B$ to 0.5 $\mu_B$\cite{picket,igor_review1,yildirim_prl1,philip}.
In section V, we will discuss this high sensitivity of the Fe-moment to the As-z position in detail in the context of giant
magneto-elastic couping since z-position and c-axis of the crystal are coupled and have to be treated equally.

In order to gain a better insight into the nature of the magnetic interactions present in Fe-square
lattice of the Fe-pnictide system, we map the calculated total energies of the  F, AF1 and AF2
configurations shown in Fig.~4a to a simple Heisenberg like model $H= E_P+ \sum_{i,j} J_{i,j} M_{i} M_{j} $
for a given fixed Fe moment $M_{i}$. For fully localized spin-systems this is a perfect thing to do but
for the case of Fe-pnictide this is only an approximation. Nevertheless, the calculated $J$s should be a good
indication of the magnetic interactions present in the system.
 We also note that these interactions are valid at  high temperatures 
above the magnetic ordering transition where spin-flips are the relevant magnetic excitations
 (and not the spin-waves).
Fig.~4b shows the {\it effective} $J_1$ and $J_2$
obtained from the energies of the F, AF1 and AF2 at given magnetic moment. The dependence of the $J$s on the magnetic moment
further suggests that the simple Heisenberg Hamiltonian is not a good model for this system. 
Here we use the term {\it effective} 
because the calculated $J_{i,j}$ is actually the interaction between spins $i$ and $j$ 
plus the infinite sum of interactions between their periodic images. 
The calculations of magnetic interactions up to
4$\it{rd}$-nearest neighbor will be discussed in the next section. 
From Fig.~4, it is clear that both $J_1$
and $J_2$ are quite large and positive (i.e., antiferromagnetic). $J_2$ is always larger than $J_1/2$
and therefore AF2 structure is the only ground state for any given moment of the Fe ion. By looking at the
exchange paths for $J_1$ and $J_2$ (shown in insets to Fig.~4), we notice that  the Fe-As-Fe angle
is around $75^{\rm o}$ and $120^{\rm o}$ for nn and nnn Fe-pairs, respectively. Hence it makes sense that
the 2nd nn exchange interaction is as strong as the nn exchange because the angle is closer to the optimum
value of $180^{\rm o}$.  

We  note that there are now a large number of studies of exchange interactions
in Fe-pnictides based on vary different methods such as mapping energies to a 
Heisenberg model\cite{yildirim_prl1,ma,wojde}, linear response theories\cite{picket,picket2,antropov}, 
strong coupling perturbation calculations  of superexchange interactions within a tight binding
model\cite{t2U}  and Kugel-Khomskii  type effective Hamiltonian with spin and 
orbital degrees of freedom\cite{kugel}. 
All these studies indicate that the major exchange interactions in Fe-pnictides are
$J_1$  and $J_2$ that are large, comparable in magnitude and antiferromagnetic (i.e. frustrated). 
This seems  to be the intrinsic property of FeAs plane that is common to all Fe-pnictide superconductors.
It is quite surprising and also very interesting that there are strong and competing antiferromagnetic
interactions in the Fe-pnictide system that  result in  a totally frustrated AF2 spin configuration. 
This is very similar to the magnetic ground state of the cuprates where the 
AF ordered 2D square lattices of the adjacent planes are frustrated\cite{yildirim_bct}.
Even though electron doping seems to destroy the long-range magnetic order, the
short range spin fluctuations will be always present and probably play an important role in the
superconducting phase, much like the high T$_c$ cuprates. 

Finally we note that there have been several neutron scattering 
measurements\cite{mag_ex_ca122,mag_j_ca122,srfe2as2_j1j2,bafe2as2_j1j2,jeff_review} of the spin-wave spectrum in 
122 systems indicating $J_1+2J_2 \approx 100 \pm 20$ meV. From Fig.~4, we get $J_1+2J_2 \approx 80$ meV 
for S=1 in 122 systems. 
Noting that our calculations were done much before the experimental measurements, the agreement is quite good and give 
confidence that DFT calculations actually work for predicting properties of Fe-pnictides. Similarly, based on our
calculations\cite{yildirim_prl1}, we had also predicted that orthorhombic lattice parameter
along the parallel-aligned spin-direction should be shorter than the axis along the anti-parallel
spin direction in the SDW phase,  which has now been confirmed by experiments\cite{jeff_review}. 
This further assures that all-electron
DFT calculations capture many fine details of the physics in Fe-pnictides.

\section{Competing Magnetic Interactions in Fe-Pnictides}

In previous section we estimated some effective exchange interactions 
from the total energies of F, AF1, AF2 but from those estimates it is not clear if the calculated
parameters are limited to the nearest neighbor interactions or not. If the
exchange interactions originate from Fermi-surface nesting then they should be long range. If 
they originate from Fe-As-Fe superexchange then they should be short range. Hence by calculating 
$J_{i,j}(R)$ as a function of
Fe-Fe distance R, we can determine if the Fermi surface nesting is the major factor in the magnetic
exchange interactions and learn more about nature of these interactions and the way they couple to
the lattice and the structural phase transition.

Here we address this issue by calculating exchange parameters from 
two different methods. 
We developed a systematic approach where the exchange parameter 
between spin-$i$ and spin-$j$ is
obtained from the total energies of a reference
magnetic configuration and those configurations obtained by  flipping
the spins $i$ and $j$ one at a time and simultaneous 
flipping both spins. From these four energies, it is possible to
obtain the exchange constant between spin $i$ and $j$. The details of this technique are presented 
in Appendix A. From now on, we refer to this method as "direct spin-flip method". Note that this
method is more appropriate at very high temperatures where the relevant spin-excitations are spin-flipping
rather than spin-waves.

\begin{figure}
\includegraphics[width=6cm]{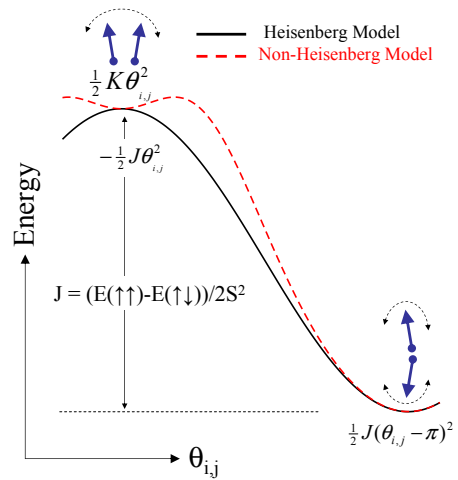}
\caption{
(color online)
Schematic representation of a spin-interaction energy versus the angle between spins $i$ and $j$ for
a Heisenberg (black) and non-Heisenberg model (red). The
linear response theory gives the second derivative of the energy  at the local spin-configuration while
the spin-flip method defines the exchange constant as
the energy difference between parallel and antiparallel spin-configurations. 
For the Heisenberg model (black) both methods give the same energy, i.e. $J$. However for non-Heisenberg model,
the linear response theory gives $K$ (i.e. ferromagnetic for this example)
while spin-flip method gives $J$ (i.e. antiferromagnetic). 
 Needless to say, both are correct! $K$ describes the spin-dynamics
near the bottom of the spin-interaction potential (i.e. spin-waves) 
while $J$ describes the spin-dynamics near the energy barrier
(i.e spins are flipping as in the paramagnetic phase). 
 }
\label{fig5}
\end{figure}

The second method is based on linear response perturbation theory using
Green function  approach within 
rigid-spin-approximation\cite{j_perp1,j_perp2,bcc_fe}
as implemented in openmx package\cite{openmx}. Basically, we
calculate the small energy change due to an infinitesimal rotation of spin $i$ and spin $j$ using Green function 
perturbation theory with the assumption that the
magnitude of spins are  fixed and only their orientations are changed. This is a
questionable approximation for Fe-pnictides because the Fe-spin magnitude is very sensitive to the
spin pattern as well as the As-position. In Sec. II, we have already seen that the spin-magnitude
goes to zero if they are forced to be aligned ferromagnetically. With this in mind, we still think
that when the system is at very low temperature where the spin-moment is fixed and spin-waves are
valid, this perturbation approach should give physically correct results. Similar approach have been
successfully used to study exchange interactions and spin-wave 
stiffness constants in transition metals such as bcc Fe\cite{bcc_fe}.  

It is very important to note that the calculated quantities from 
spin-flip and linear-response-theory are not the same thing and therefore these two methods can
give totally different results for non-Heisenberg spin-interactions. This is schematically shown
in Fig.~5. The linear response-theory gives the second derivative of the spin-interactions from infinitesimal
rotation of the spins near their equilibrium positions. On the other hand, the spin-flip method gives the energy
difference between parallel and antiparallel spin configurations. When we deal with Heisenberg interaction, $J(\theta)$
is proportional to $\cos(\theta)$ and therefore both the second derivative and the energy difference give the same answer.
However for a non-Heisenberg model where the dependence of $J$ on the angle is not a simple cosine as shown
in Fig.~5, the linear response theory will give the 2nd derivative at the local structure, i.e.  $K$ in Fig.~5 while the
spin-flip method will give $J$ and therefore the two methods will differ. This should not be considered as one of the method
is not working but rather as an evidence that the spin-spin interaction is not a traditional Heisenberg type. As we shall
see below, this is indeed the case for Fe-pnictide where for the parallel-spin direction in the stripe phase, linear
response theory will give a ferromagnetic interaction while the spin-flip method will give an antiferromagnetic interaction.
Since the linear response theory probes the energy changes near the local-spin configurations due to small spin-rotations,
the $Js$ from this method should be used to explain the experimental spin-wave spectrum. However for spin-dynamics near or
above the paramagnetic phase transition the spin-waves are not valid and therefore the $Js$ from spin-flip method 
is more appropriate. Of course, since the $Js$ depend on the spin-configurations considered one can not study the phase
diagram of Fe-pnictides from zero to high temperatures with a single Heisenberg like Hamiltonian. In that case, it is important
that one goes beyond the Heisenberg picture and consider more complete models such as 
Kugel-Khomskii   Hamiltonian where the orbital and spin-degrees of freedom are 
treated self consistently on equal footing\cite{kugel}.

\begin{figure}
\includegraphics[width=6cm]{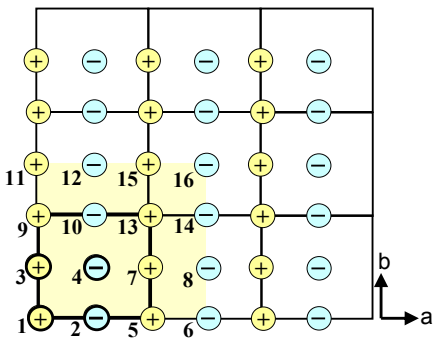}
\caption{
(color online)
The $3\sqrt(2)\times3\sqrt(2)\times1$  supercell considered in calculations of the
exchange parameters in real-space. For simplicity,
we show only the Fe ions with their number labels and 
spins (+ up, - down) according to AF2  ordering. 
The J$_{i,j}$ for each iron pairs $i$ and $j$ are
 plotted in Fig.~6. The shaded region indicates the half of the supercell in which  we can extract the
 exchange interactions between any pairs of spins (which are numbered from 1 to 16 for convenience).
 }
\label{fig6}
\end{figure}

Finally, we note that both methods discussed above are based on periodic supercell approach and 
therefore what we calculate is actually the sum of the interactions between spin $i$ and $j$ and the interaction
between their periodic images. Hence it is important to consider a very large supercell to make sure
what we get is actually the individual exchange constant between spin $i$ and spin $j$. Hence, we
consider $3\sqrt(2)\times3\sqrt(2)\times1$ cell which has total 144 atoms, 36 of which are iron. 
 Fig.~6 shows the labeling of the 32 iron atoms in the large supercell. The direct spin-flip
calculations were done using the plane-wave code pwscf\cite{pwscf} with cutoff energy of 30 Ry and charge cutoff
of 240 Ry using PBE-GGA exchange functional. We used 2x2x3 k-points. The perturbation calculations with
rigid-spin-approximation were carried out using the package openmx\cite{openmx} which implements numerical atomic
arbitals and norm-conserving pseudopotentials. We used equivalent cutoff, k-point grid and the same PBE
exchange functional as the pwscf calculations. We used experimental atomic positions for the LaOFeAs system
 without any  structural relaxation. Both methods (i.e. pwscf and openmx) give an iron magnetic moment close to 
 $\approx 2 \mu_B$, a typical value  obtained from pseudo-potential based methods. For convenience, in the discussion below,
we report  $J S^2$ by setting S=1 (rather than taking S=2 and recalculating $Js$).

\begin{figure}
\includegraphics[width=8cm]{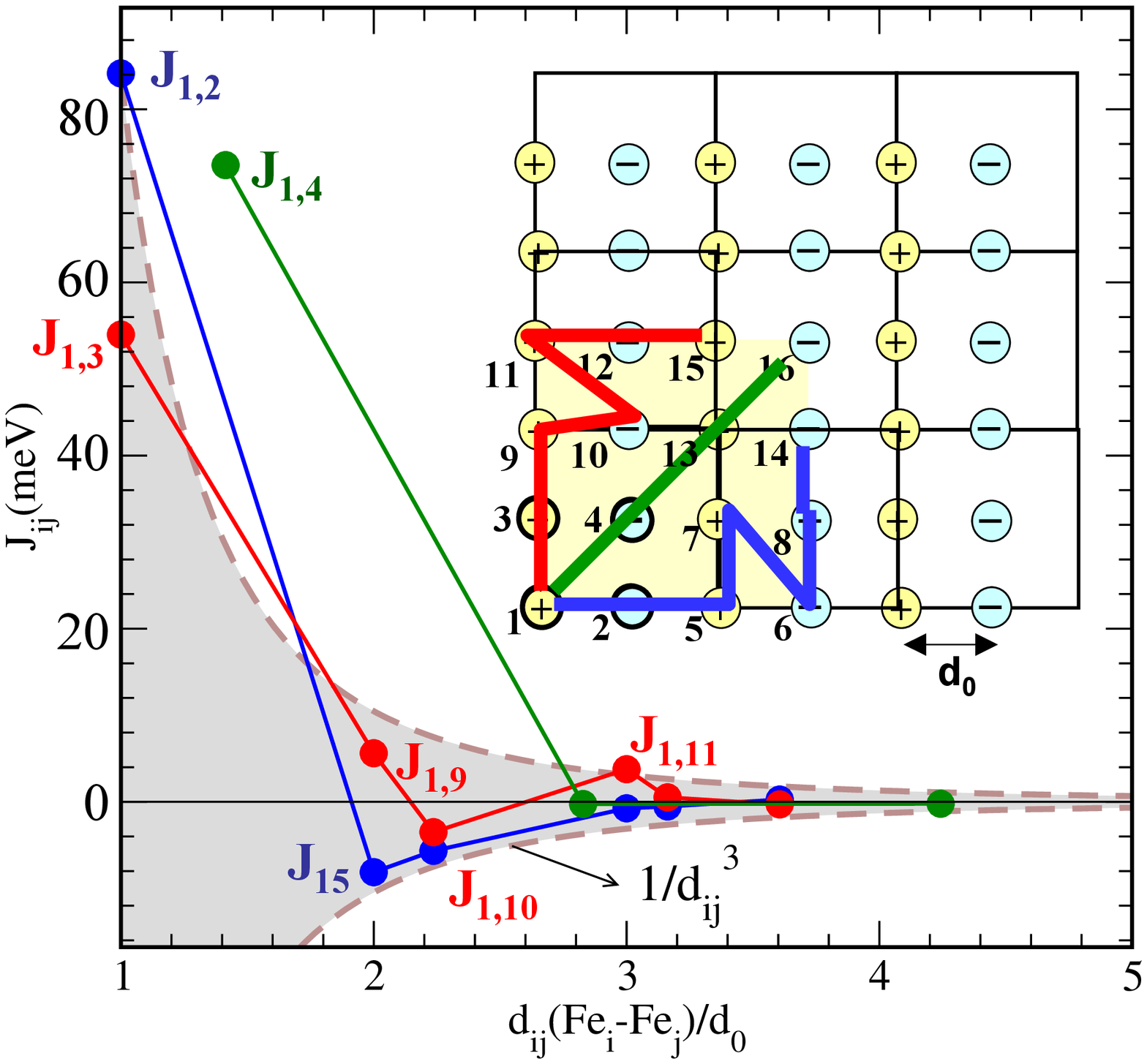} \\
\includegraphics[width=8cm]{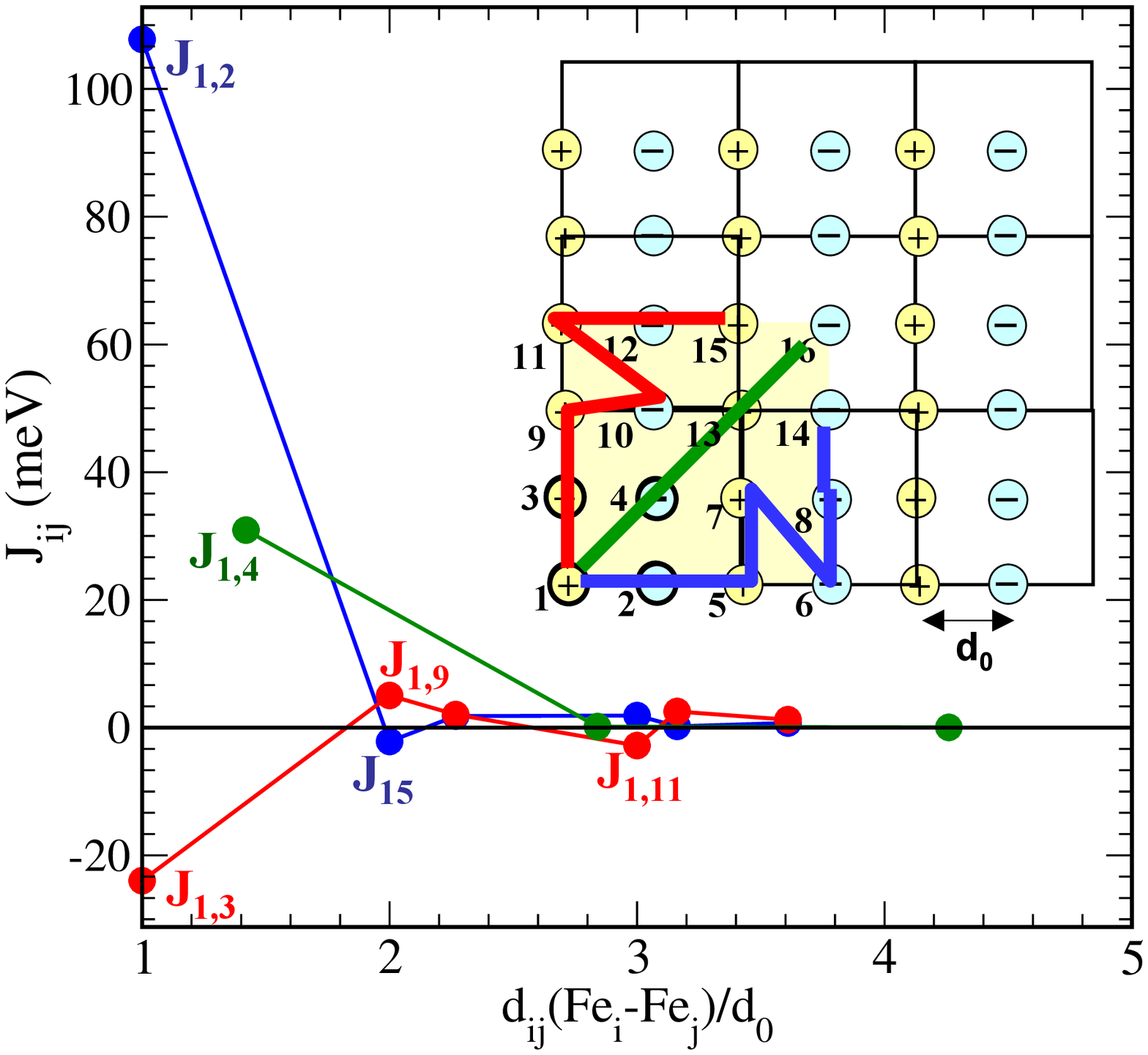} \\
\caption{
(color online)
The calculated exchange constants J$_{i,j}$(R)  ( $S$  in $J S^2$ is taken to be 1)
along three different paths as shown in the inset. Top panel shows the results 
 from  direct spin-flip method while the bottom panel shows results
from linear response perturbation method within rigid-spin-approximation. 
Both methods give strong and antiferromagnetic   short range
interaction along the diagonal direction   while they differ for the
parallel aligned spin-direction  (i.e. b-axis).  The dashed curves and the shaded area in top panel shows
that the exchange interactions along  the $a$ and $b$ axes are oscillatory  with an 
envelop decaying as  $1/R^3$, much like in the bcc-Fe\cite{bcc_fe}.  
 }
\label{fig7}
\end{figure}

Our results,  for the exchange constants, $J_{i,j}(R)$ 
from both spin-flip and perturbation methods are  shown in Fig.~7.  
We first discuss top panel in Fig.~7 which is from
direct spin-flip method. The strongest  interactions, i.e. $J_{1,2}, J_{1,3}, $ 
and $J_{1,4}$ are all  antiferromagnetic and comparable to each other, consistent
 with our previous results obtained from the total energies of three spin configurations 
 discussed in section II. We note that the symmetry
between $a$ and $b$ axis is broken and we obtain different $J_1^a$  (i.e. $J_{1,2}$) and
$J_1^b$ (i.e. $J_{1,3}$). However the difference is small and they are both antiferromagnetic.
The $J_2/J_1$ is around 1 (here $J_2$ is $J_{1,4}$) and therefore the AF2 (i.e. SDW) is the
ground state. Fig.~7 also shows how the exchange constants decay with  distance along the Fe-Fe square
diagonal, and along a- and b-directions. We note that the anti-parallel-spin alignment   is taken to
be along the a-axis. Interestingly,  $J_{i,j}(R)$ already changes sign and become ferromagnetic 
at the 2nd (i.e. $J_{1,5}$)  and 3rd shells (i.e. $J_{1,10}$) along the $a-$ and $b$-axes, respectively.
As shown in Fig.~7, along the $a-$  and $b-$directions, $J_{i,j}(R)$ has an oscillatory
character with an envelop decaying as $1/R^3$. One exception to this decay rate is the $J_{1,4}$ (i.e. $J_2$), 
the exchange interaction along the Fe-Fe square diagonal direction. $J_{1,4}$ is very large (i.e. way above the 
$1/R^3$-envelop) and then basically goes to zero for further distances (i.e. $J_{1,13} = J_{1,16} \approx 0$).
This suggest that the origin of   $J_{1,4}$  is probably not the 
Fermi-surface nesting or other long-range exchange interactions but rather 
local superexchange interactions through As-p orbitals. In contrast, the
nature of the exchange constants along the $a-$ and $b-$directions  are very different.
They are long range and decay with  $1/R^3$, much like in  bcc-Fe\cite{bcc_fe}.

The bottom panel in Fig.~7 shows the  calculated exchange parameters $J_{i,j}(R)$ obtained from
linear response perturbation approach with rigid-spin approximation. Similar to results from
spin-flip method, the
exchange interaction along the Fe-Fe square is antiferromagnetic and very short range while along the
$a-$ and $b$-directions it has oscillatory character with sign change and slow decay.
The most important difference between the spin-flip and perturbation methods is the nearest-neighbor exchange
interaction along the stripe direction, i.e. $J_{1,3}$.
 While spin-flip method gives this interactions as antiferromagnetic
the linear response theory suggests it is weak but ferromagnetic. 
As discussed above and shown in Fig.~5, this 
difference between two methods is a nice evidence that 
the spin-interactions along the stripe direction is 
probably not a classical
Heisenberg.  In conclusion, 
due to slow $1/R^3$ decay of $Js$ along the stripe direction and its
different sign obtained from spin-flip and linear response methods
strongly suggest that the spin-interactions along the stripe
direction is unusual, long-range and probably  related to Fermi-surface nesting.

We note that our results  from perturbation theory  are in 
agreement with the earlier calculations reported from 
Pickett's group in Ref.\onlinecite{picket2}  where the spin-exchange
constants were calculated in various 122 systems using linear response theory and 
small but ferromagnetic  exchange interactions are found along the parallel-spin direction. 
We emphasize that since the 
exchange interaction along the stripe direction (i.e b-axis) is ferromagnetic, the frustration is totally 
removed and there is nothing  special for the ratio $J_1^a/2J_2$ being $ \approx 1$ anymore. Finally, we note that
there is an other interesting report by Balashchenko and Antropov\cite{antropov} where 
the exchange interactions are calculated in real space
as a function of As z-position using linear response theory. As expected, they observe magnetization changes from
$1.3 \mu_B$ to $0.31 \mu_B$ with a small displacement of arsenic z-position which in turn changes the Fe-As bond
distance by about 0.1 \AA. However the calculated exchange parameters do not agree 
with those reported from Pickett's group even though both groups use linear response theory.
They found that all the major interactions are antiferromagnetic with significant anisotropy along the parallel
and antiparrallel spin directions. This anisotropy is found to be very sensitive to the As z position. 
However for all z-positions studied, the exchange along the stripe direction is always antiferromagnetic.
 The authors also conclude that the interactions are long range in agreement with our results for the
 spin directions along the $a-$ and $b-$axis. However we emphasize that from both methods used here, we find that
 the diagonal spin interaction (i.e. $J_2$) is very strong and antiferromagnetic for only the nearest neighbor and then
 basically becomes zero for further distances.

Since two methods shown in Fig.~7  give
opposite spin-interactions along the stripe direction (i.e. parallel spin-direction), 
an interesting picture emerges from these two calculations. At   high temperature
where the system is paramagnetic or close to magnetic ordering , we find that the major magnetic interactions are
antiferromagnetic and frustrated. The exchange interactions along a- and b-directions are comparable to each other
as it should be due to tetragonal symmetry of the paramagnetic state. As the system orders and the
spin-flip excitations become more and more spin-oscillations as in the case of spin-waves, 
the band structure is modified (i.e.
the symmetry between  $d_{xz}$ and $d_{yz}$  is broken) yielding very different exchange interactions
along  a- and b- directions. In the low-temperature limit where the spins are pretty much fixed,
we expect the perturbation results valid and should replace the spin-flip results. 

Due to non-Heisenberg nature of the 
spin-interactions that we obtain here, one has to be careful in modeling these systems using a Heisenberg like model. 
For a given spin-configuration, it is probably OK to use  a Heisenberg model to study low-energy
excitations in that configuration (such as spin-waves in SDW ordered state). However if one wants to study the whole phase
diagram as a function of temperature, a simple Heisenberg model with a fixed $Js$ is clearly not appropriate. As the spins
rotate or change configurations, one will obtain totally different exchange constants. Hence in this case, one needs to go
beyond the classical Heisenberg model to treat the Fe $d-$orbitals and spin degrees of freedom on equal footing. Hence a
Kugel-Khomskii like Hamiltonian could be more appropriate for Fe-pnictide systems. Recently a complicated phase diagram of
spin-orbital ordering of Fe-pnictide has been studied in Ref.\onlinecite{kugel} and we refer the reader to this study for
details.

We finish this section with a brief discussion  of the experimental determination of
 the exchange parameters  in Fe-pnictides from inelastic neutron scattering experiments. 
Even though, Heisenberg model is not appropriate as discussed above, it is probably good
enough to describe the spin-wave excitations at low temperatures where the spins are not flipping and therefore
the exchange constants are not changing.  A minimal spin Hamiltonian for Fe-pnictide  may be 
written as
\begin{equation}
{\cal{H}} = \frac{1}{2}  \sum_{i} \sum_{\alpha=a,b,d,c}  J_{\alpha} \vec{S}_{i} \cdot \vec{S}_{i,\delta_\alpha} 
- D \sum_{i} S_{ix}^{2}
\end{equation}
where the $i$-summation  runs  over all Fe-spins and $J_\alpha$ ($\alpha=a,b,c,d$)  
 are the exchange interactions along
the antiparallel spin direction $a$, along the parallel spin-direction $b$, along  the c-axis, and  finally along
the Fe-Fe square  diagonal, respectively. The last term is the easy-axis single ion anisotropy  which originates from
spin-orbit coupling. Usually it is small but here due to very large $J_a$ and $J_d$, it is important to keep this term 
which can give large spin-gap at gamma that is proportional to $\sqrt{ D \times (J_a+2J_d)}$.
The spin-wave spectrum of this Hamiltonian can be easily obtained by considering
the AF2-spin configuration as  helimagnetic ordering with modulation wavevector $Q=(\pi,0,\pi)$ such that
the spins are ordered antiferromagnetically along the $a-$ and $c-$ axis and ferromagnetically along the $b$-axis.
We note that the unit cell of the 
helimagnetic spin-structure is twice smaller than the chemical cell along all directions (i.e. $a_M=a_o/2, b_M=b_o/2,$
and $c_M=c_o/2)$. Within this helimagnetic description of the cell, the above model Hamiltonian has  a 
single spin-wave mode
\begin{eqnarray}
\omega(\bold{q})  &= &2 S \sqrt{A_{\bold{q}}^2-B_{\bold{q}}^2} \nonumber \\
	A_\bold{q} &=&  J_a  - J_b [1-\cos(q_b)]+2J_d +J_c +D \\
	B_\bold{q} &=& J_a \cos(q_a) +2 J_d \cos(q_a)\cos(q_y) + J_c \cos(q_c) \nonumber
\end{eqnarray} 
and the zero-temperature inelastic structure factor (which is proportional to inelastic neutron scattering intensity)
\begin{equation}
S(\bold{q},\omega) = \sqrt{\frac{A_q-B_q}{A_q+B_q}} \delta(\omega-\omega_q)
\end{equation}

\begin{figure}
\includegraphics[width=8cm]{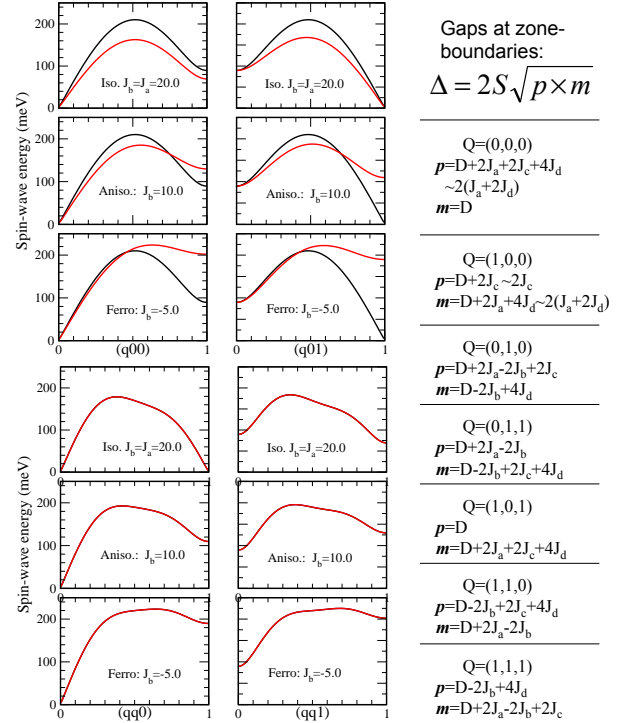}
\caption{
(color online)
The $ab-$domain-averaged spin-wave spectrum along (q,0,h) and (q,q,h) directions (h=0, 1).
The black and gray (red) curves are due to
spin-waves coming from domain-a and domain-b where the stripe direction is along the a- and b-axis, respectively.
Note that with a proper constant momentum  and/or energy scans, it should be possible to resolve two modes 
from two domains
and uniquely determine the sign of exchange parameter J$_b$. 
On the right, we give the analytical expressions for the
spin-wave gaps at different Brillouin zone boundaries. The plots are 
obtained using the following values (all in meV);
$J_d=40.0,J_a=20,J_c=5.0, D=0.015$.
 }
\label{fig8}
\end{figure}

After having determined the eigen-spectrum of our model Hamiltonian, we now discuss how one may measure them
using inelastic neutron scatting. Because of the tetragonal symmetry of the paramagnetic phase, 
during the SDW transition the
single crystal samples will probably have orthorhombic twinning which we call ab-domains. 
In one domain the stripe direction (i.e. parallel aligned
spin direction) will be along the a-axis while in the other domain it will be along the b-axis. 
In other words, even though we have a
single crystal sample, we can not distinguish the wave-vector ${\bf q}=(q_a,q_b,q_c)$ 
from ${\bf q}=(q_b,q_a,q_c)$. Hence, from
inelastic neutron scatting one will therefore probe the superposition of the spin-wave modes 
at these two wave-vectors, i.e. $\omega(q_a,q_b,q_c)$  + $\omega(q_b,q_a,q_c)$. As we shall see below, due to 
very different ordering of spins along the $a-$ and $b-$ axes, the superposition of these two modes
can be, in principle,  resolved at high energies and in particular at zone-boundaries.

In Fig.~8 we show the $ab-$domain averaged spin-wave spectrum, i.e. $\omega(q_a,q_b,q_c) + \omega(q_b,q_a,q_c)$, along
various directions. For the exchange parameters, we consider three cases; isotropic model ($J_a = J_b$),
anisotropic model (i.e. $J_b = J_a/2$), and finally ferromagnetic model ($J_b < 0 $). The values of the 
exchange parameters used in the calculations are given in Fig.~8. From Fig.~8, it is clear that contributions due to 
$a-$domain (red-curve) and $b-$ domain (black curve) are quite different and could be resolved experimentally. 
Interestingly for the isotropic model ($J_a = J_b$), the spin-wave spectrum is the most anisotropic near the 
zone-center and near the middle of the $(1/2,0,0)$ as shown in Fig.~8. This is due to the fact that the spins are aligned 
antiparallel and parallel along $a$ and $b$  axes, respectively while their interactions are forced to be equal 
and antiferromagnetic. This gives very different dispersion along the $a$- and $b-$axis and therefore one can 
resolve the two peaks in 
the $ab-$domain averaged spectrum. 
From Fig.~8, it is clear that near the $(1,0,0)$,  the isotropic model have two modes near zero energy 
  while the anisotropic and ferromagnetic models have one mode that has a
  huge spin-wave gap that is proportional to $\sqrt{(J_a-J_b)( 2J_d-J_b)}$. The analytical expressions for
  the spin gaps at various Brillouin zone boundaries are given in Fig.~8. Finally, we note that
  $(qq0)$ direction is a special one where we should see only a single mode which has a tiny gap at (100) 
   if the 
  interactions are isotropic and a gap about the half of the maximum spin-wave energy for the anisotropic model.
  For the ferromagnetic model, the gap is almost the same as the maximum spin-wave energy (see left bottom panel 
  in Fig.~8).

\begin{figure}
\includegraphics[width=7cm]{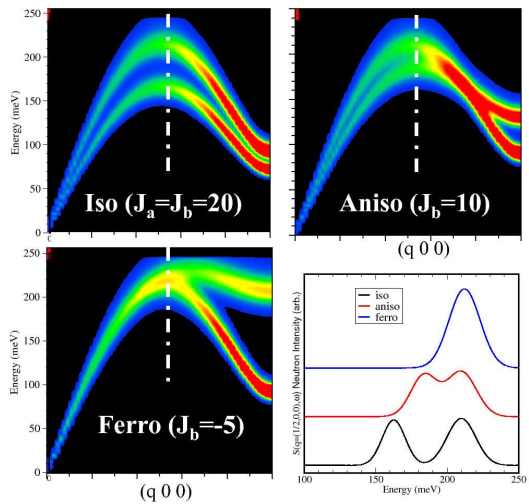}
\caption{
(color online)
The contour plot of calculated $ S({\bf q},\omega)$ for iso, aniso, and ferro-model for $J_b$ along 
(q,0,0)-direction. The energy scans at constant q=(1/2,0,0) for the three models are indicated by dashed-line
and shown in the bottom right panel. Note that despite  the presence of orthorhombic-twinning, 
the spin-wave spectrum of the three
models are quite distinct suggesting that one may uniquely determine the sign of $J_b$. All these are based on
the assumption that the Stoner-continuum will not over-damp the spin-wave spectrum.
 }
\label{fig9}
\end{figure}

In Fig.~9, we show the $ab-$domain averaged inelastic-structure factor to get an idea about the 
intensities of the modes from two domains. The calculated spectrum is convoluted with Gaussian and
 $5\%$ energy resolution. Fortunately the intensities from both domains are comparable
and therefore it should be feasible to detect these modes in a multi-domain sample. As an example, in Fig.~9, we show 
energy scans at constant wavevector $Q=(1/2,0,0) $ for the three modes, which show very distinct spectrum for each model.

As a final note, we point out that  in practice, the observation of spin-waves in Fe-pnictide
could be problematic at high energies due to strong spin-wave damping by Stoner-continuum. 
However very recent inelastic neutron scattering measurements on Ca122 system\cite{mag_ex_ca122} has 
revealed steeply dispersive and well-defined spin waves up to an energy of $\approx 100$ meV. 
Unfortunately the resolution and the quality of these recent data are not good enough to carry out a detailed 
analysis as discussed here in order to determine the sign of $J_1^b$. 
We hope that in the near future there
will be more  experimental data with better resolution and quality.

\section{ Structural Phase Transition  in  Fe-Pnictides}

We next discuss the implication of the magnetically frustrated AF2  configuration on the
structural distortion which is shown to be a common feature of 1111 and 112 parent compounds\cite{jeff_review}. 
Experimentally it has been demonstrated that magnetic SDW ordering and the tetragonal-orthorhombic 
distortion are closely coupled\cite{pt_order}. Interestingly for 1111 systems (i.e. LaOFeAs)\cite{cruz} , 
the tetragonal distortion first takes place about 20-30 K higher in temperature than 
the magnetic transition.  The transition for 1111 system is somewhat weak. 
On the other hand, for the 122 systems (such as BaFe$_2$As$_2$), it is clear that both transitions occur at 
the same time. They are more or less strong first-order type and one study suggests that the
structural distortion is proportional to the magnetic order parameter (rather than its square)\cite{pt_order}.

\begin{figure}
\includegraphics[width=8cm]{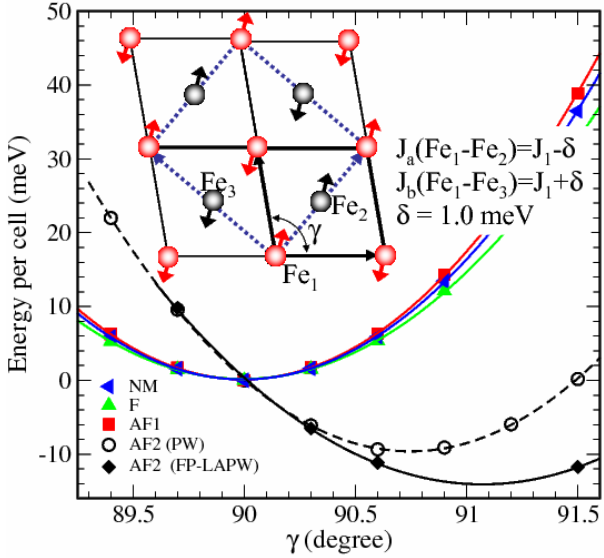}
\caption{
(color online)
The total energy per cell versus the angle $\gamma$ for non-magnetic (NM), Ferromagnetic (F) and two
antiferromagnetic (AF1 and AF2) spin-configurations. Note that only the AF2 spin configuration
yields structural distortion. The inset shows that as $\gamma$  increases,  the ferromagnetic aligned
Fe ions (i.e., Fe$_1$-Fe$_2$) get closer while the antiferromagnetically aligned ions (i.e., Fe$_1$-Fe$_3$) move
apart, breaking the four-fold symmetry and thus the degeneracy of the d$_{xz}$ and d$_{yz}$ orbitals.
For the AF2, the solid and dashed lines are from pseudo potential plane wave (PW) and FP-LAPW
calculations, respectively.
 }
\label{fig10}
\end{figure}

In order to establish a connection between the structural distortion and the magnetic ordering,
we calculate the total energy as a
function of the $\gamma$ angle 
as shown in the inset to Fig.~10 for NM, F, AF1, and AF2 spin configurations (see Fig.3).
When $\gamma=90^{\rm o}$, we have the original tetragonal cell. Once the
$\gamma $ deviates from $90^{\rm o}$, the original $\sqrt{2}\times\sqrt{2}$ structure (shown as dashed
line) is no longer tetragonal but orthorhombic (i.e., the cell length along a and b axes are no longer
equal).
We note that for $\gamma=90^{\rm o}$, the orbitals $d_{xz}$ and
$d_{yz}$ are degenerate and therefore one may think that the system is subject to  symmetry lowering for reasons
similar to those in a Jahn-Teller distortion. However as shown in Fig.~10, we do not see any distortion for any of 
the NM, F, and AF configurations. 

The total energy versus $\gamma$ plot shown in Fig.~10 clearly indicates that only AF2 ordering 
distort the structure  with  $\gamma = 91.0^{\rm o}$, which is in good agreement with the experimental 
value of $90.3^{\rm o}$. The FP-LAPW method with LDA approximation we get M=0.48 $\mu_B$ which is in
excellent agreement with the experimental value of $0.35 $ $\mu_B$. Hence all electron-calculation is able
to reproduce both the observed structural distortion and the small Fe-moment simultaneously. On the other hand,
PW calculations with GGA approximation (dotted line) give good structural parameters when spin-polarization is 
allowed but then the 
calculated moment is too large. Somehow the magnetic ground state is over-stabilized in PW calculations. Therefore one
has  to be careful in studying the magneto-elastic 
couplings by PW methods, which will be underestimated. 
The net energy gain by
the structural distortion shown in Fig~10 is about 12 meV per cell, which is of the same order as the temperature at
which this phase transition occurs.  We also considered two types of
AF2 where the spins along the short axis are aligned parallel or anti-parallel (see Fig. 3c-d). 
These two configuration
are no longer equivalent. According to our calculations the configuration in which the spins are ordered parallel 
along the short-axis is the ground state. This prediction\cite{yildirim_prl1} has now been confirmed by neutron scattering 
measurements for several 122  systems\cite{jeff_review}, giving us confidence that first-principles 
calculations describe many fine
details of Fe-pnictide systems accurately.

Even though full structural optimization discussed above clearly shows that AF2 ordering gives rise to 
observed orthorhombic distortion with the  long axis along the anti-parallel spin direction,
it is not obvious why this happens. Naively one would expect the opposite, i.e. 
parallel spin-direction is the frustrated one and therefore it should get longer to decrease the 
antiferromagnetic interactions. This would be the case if we had Cu-O-Cu linear bond. 
However in the Fe-pnictide case, the
arsenic ions are not directly between two Fe-ions (see Fig.~2)
but they are above/below the Fe-squares. Hence when Fe-Fe distance is increased 
along a-axis, the Fe-As-Fe bond angle also increase along this direction 
while Fe-As distance does not change at first order. 
In fact, As ions are pulled down to keep the rigid FeAs bond fix and increase the angle.
However the net effect of the increase in angle on the exchange interaction itself 
is not easy to predict due to
complicated nature of the problem.

To determine if the   exchange interaction $J_1^a$ increases   while $J_1^b$
decreases  during the orthorhombic transition,  we calculate the
lattice dependence of the exchange parameters, i.e. $dJ/da$,  and  then
 develop   a simple spin-Peierles like model which successfully 
 explains both the transition and the lattice parameters of F, AF1 and AF2
 spin configurations within a unified picture.

\begin{figure}
\includegraphics[width=8cm]{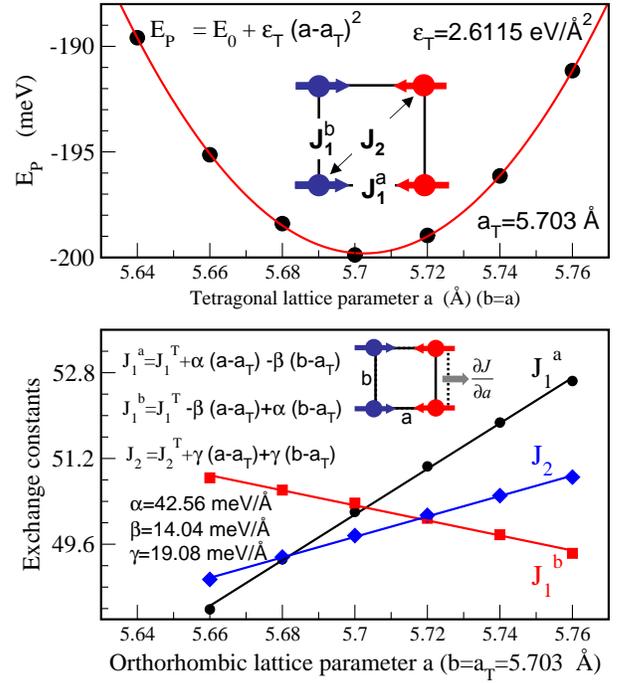}
\caption{
(color online)
Top: The paramagnetic portion of total energy ($E_P$) versus the tetragonal lattice parameter a 
(a and b are taken to be equal). 
The inset shows the SDW ordering, the relevant exchange constants and the fit to the total
energy.
Bottom: The linear expansion of the exchange constants with respect to lattice deformation. The
tetragonal cell is distorted along a-direction while the b-axis is kept constant at $b=a_T=5.703$ \AA.
Points are the actual calculations and solid lines are the linear fit.
 }
\label{fig11}
\end{figure}

  In order to obtain the lattice-parameter dependence of
the $Js$, we calculate the energies of F, AF1 and AF2 configurations for different  tetragonal lattice
parameters and then extract the E$_P$, the paramagnetic portion of the total energy as a function
of tetragonal cell $a$. This is shown in the top-panel of Fig.~11. The tetragonal elastic constant is
then extracted by fitting the total energy to a quadratic form as shown in the figure. After having
determined the paramagnetic contribution of the total energy, we then apply tetragonal to
orthorhombic distortion by varying the lattice parameter along a-direction. 
For each distortion, we then calculate the energies of F, AF1 and AF2
spin-configurations. For each spin-configuration  the internal atomic coordinates such as 
As z-position  are always optimized. Using the energy expressions given in Fig.~3, we then extract the $J_1$
and $J_2$ which are given in the bottom panel of Fig.~11. We note that $J_1^a$ increases and 
$J_1^b$ decreases  linearly as the a-axis is elongated. Similarly, the $J_2$ always increase with increasing lattice
parameters near the paramagnetic optimum tetragonal cell parameter $a_T$ as shown in Fig.~11. 
Hence it seems that both $J_1$ and $J_2$ 
increases with increasing bond-angle.

Now, using the lattice parameters dependence of $J_1^a, J_1^b,$ and $J_2$ calculated in Fig.~11, one can easily 
obtain the new lattice parameters of the magnetic cell for ferro, AF1 and AF2 spin
configurations which are summarized below:

\begin{eqnarray}
F \rightarrow a=b&=&a_T 
- (\frac{2\delta}{\epsilon_0} +\frac{\alpha-\beta}{\epsilon_0} )<S^2>  \nonumber \\
&=& 5.677 \AA \; (5.625 \AA) \\
AF1 \rightarrow a=b&=&a_T 
- (\frac{2\delta}{\epsilon_0} -\frac{\alpha-\beta}{\epsilon_0} )<S^2>  \nonumber \\
&=& 5.699 \AA \; (5.701 \AA) \\
AF2 \rightarrow a&=&a_T 
+ (\frac{2\delta}{\epsilon_0} +\frac{\alpha+\beta}{\epsilon_0} )<S^2>  \nonumber \\
&=& 5.739 \AA \; (5.734 \AA) \\
b&=&a_T 
+ (\frac{2\delta}{\epsilon_0} -\frac{\alpha+\beta}{\epsilon_0} )<S^2>  \nonumber \\
&=& 5.696 \AA \; (5.668 \AA) \
\end{eqnarray}
Here the numbers given in parentheses are the results from self consistent full cell
relaxation calculations. From above results, it is clear that this simple model
 explains most of the observed features such as
F and AF1 ordering does not distort the lattice. Ferro magnetic cell has the smallest lattice parameter because
the system wants to make both $J_1$ and $J_2$ smaller as they are antiferromagnetic and we are forcing
the spins ferromagnetically ordered. Similarly, for AF1 ordering, we get competitions between 
J$_1$ which wants to increase the lattice while the $J_2$ term is still not satisfied and therefore
it wants the lattice shrinks, yielding a lattice parameter larger than ferro but smaller than AF2. In
the case of AF2, the J$_2$ is totally happy and want to increase the lattice. Our model for AF2
nicely predicts orthorhombic distortion with right lattice parameters.

\begin{figure}
\includegraphics[width=7.5cm]{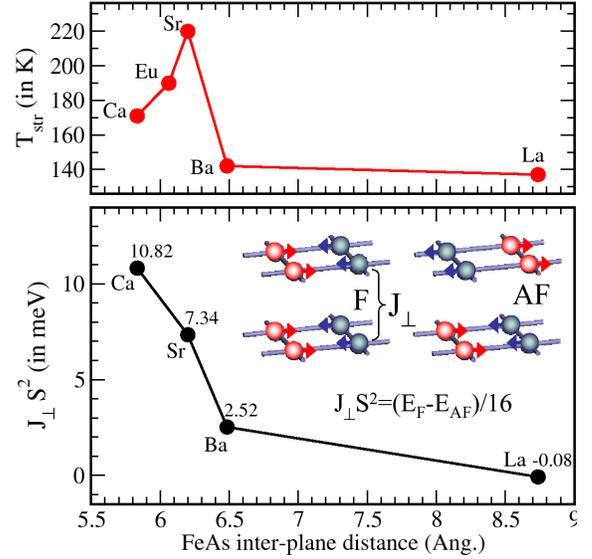} 
\caption{
(color online)
Top: The structural phase transition temperature $T_{str}$ versus
FeAs interplane distance for M122 (M=Ca, Eu, Sr, and Ba) and La1111 systems.
Bottom: The calculated interplane exchange interactions $J_\perp$ (in meV)
versus FeAs interplane distance. For the La1111 system, the energy difference
between AF and F magnetic configurations (shown in the inset) is too small to
get an accurate number but it is less than 0.2 meV. 
Experimentally\cite{jeff_review} $J_\perp$ in 1111 systems is also found to be
very weak with both positive and negative sign depending on the
rare-earth in the system.
The inset shows the
ferro and antiferro alignments of the FeAs planes in the unit cell. Here 
$E_F$ and $E_{AF}$ are the total energies per cell (i.e. 8 Fe atoms).
Note that while $J_\perp$ drops sharply with increasing Fe-Fe interplane distance, 
$T_{str}$ first increases and then drops with increasing Fe-Fe distance.
}
\label{fig12}
\end{figure}

Finally we note that all of our discussion given above is basically based on a single FeAs layer without inter-plane 
interactions. In reality these systems order three dimensionally at the structural phase transition.  
This raises a questions;  is there any correlation between the structural phase transition $T_{str}$ and 
the inter-plane magnetic interaction $J_\perp$ between FeAs planes. 
 Normally the larger the
$J_\perp$ the higher the $T_{str}$ should be.  To answer this question, we have 
calculated $J_{\perp}$ for different 122 systems and plot it as a function of Fe-Fe plane distance in Fig.~12.
From this figure it is clear that structural phase transition temperature
 $T_{str}$ does not have a monotone dependence on Fe-Fe
distance while $J_\perp$ does. This suggest that there must be an other factor which effects the $T_{str}$ in
 an opposite way. One possibility could be the shear modulus of the system. If the c-axis is short, then the chemical 
bonding between FeAs planes would be stronger making  the structural distortion difficult. On the other hand, 
the smaller c-axis
means the larger $J_\perp$ which means larger in-plane AF2 correlation that  drives the structural phase
transition. Since the effects of shear-modulus and J$_\perp$ are opposite, 
 T$_{str}$ could have a maximum at a particular 
FeAs interplane  distance as indicated by the experimental data.
Currently we are developing a phenomenological Landau-theory along these lines
 to address the remaining issues discussed above and our results
 will be published elsewhere\cite{harris_taner_feas}.
 We note that there have been already a large number 
 of theories\cite{pt_th1,pt_th2,pt_th3,pt_th4} to describe this coupled 
 structural and magnetic phase transitions.  
 We refer the reader to these studies for
 details\cite{pt_th1,pt_th2,pt_th3,pt_th4}.

\section{Giant Magneto-Elastic Coupling in Fe-Pnictides}

In order to get a general understanding giant magneto-elastic coupling present in  iron-pnictides, 
here we consider one example
of each class of pnictides; namely CaFe$_2$As$_2$ for the 122 system with the smallest Ca-ion available and
the BaFe2As2 with the largest metal Ba. For the 1111 system, we consider LaOFeAs. 
We also study a doped
122-system, i.e. Na$_{0.5}$Ca$_{0.5}$Fe$_2$As$_2$. Since in our 
$\sqrt{2}\times\sqrt{2}$-cell we have
four chemical formula, we consider a supercell where two Na and two Ca are ordered. For each given
system, we have performed full structural optimization  including the lattice
parameters and the atomic positions. We consider our optimization is converged when the maximum force on
each atom is less than 0.005 eV/\AA~ and the pressure is less than 0.1 kbar.
We have performed the full structural optimization for non-magnetic (NM), i.e. "non-spin polarized",
checkerboard antiferromagnetic (AF1) and stripe (AF2) spin configurations. Our results are summarized
in Table~1.  As expected, the ground state for all four systems is the stripe AF2 phase and the
optimized parameters are in good agreement with the experimental data at ambient conditions.

The most striking and surprising finding listed in Table~1 is the giant dependence of the
optimized c-lattice parameter on the spin-configuration considered. For the case of CaFe$_2$As$_2$,
we note that AF1 configuration is the next stable  state (after the AF2) but the c-value is
significantly reduced; 11.63 \AA~ versus 10.60 \AA~ for AF2 and AF1 spin configurations, respectively.
This difference in c is not due to different AF1 and AF2 spin configuration 
but due to different Fe-spin state in AF1 and in AF2.
The difference is even larger, when the Fe-magnetism is totally ignored (i.e. non-spin polarized calculations).
The optimized z-value for NM-state is 10.39 \AA, which is 1.34 \AA~  shorter than the experimental value
at ambient pressure.  We note that the optimized  lattice parameters, a=5.65 \AA~ and c=10.39\AA~ for the 
NM phase are in reasonable agreement with the neutron data in  the 
collapse phase (a=5.8 \AA~ and c=10.6 \AA)\cite{cTphase}. 
Hence from these results and 
the recent experimental observation of the collapse-T phase, one can reach the conclusion that
the Fe-moment should be present in 122 systems at all times at ambient pressure. 
This is because we know that these systems are ordered
in an AF2 spin configuration when they are not doped or superconducting. When they are doped and
superconducting we do not see huge changes in their c-lattice parameters. This indicates that even though
the AF2 long range ordering is destroyed with doping or we are at temperatures above the magnetic
ordering  transition T$_N$, the Fe-spin should be present in the system. Otherwise, we should see the expected huge
reduction in the c-axis.  In other words, the iron-pnictide system should be considered as paramagnetic
i.e. with the on-site non-zero iron moment without any long range order. We note that the 
paramagnetism is different than the "non-magnetic" case that we consider in our DFT calculations 
where we force equal up and down spins in each orbital. The non-spin polarized
calculations  should not be considered as a model
for the paramagnetic system. With the standard density functional theory, there is no way to treat a
paramagnetic system (i.e. we have the spin-degrees of freedom at each site but no long range order).

\begin{table}
\caption{Various optimized  structural parameters
for NM, AF2, and AF1 spin configurations, respectively. Note that the c-axis from
non-spin polarized and AF1-spin configurations are  significantly smaller than the
 experimental data at ambient
conditions. The zero of energy is taken as the energy of the NM-case. 
The experimental data are taken 
from Refs[\onlinecite{cafe2as2,canaxfe2as2,bafe2as2,cruz}].
$^*$The AF1 configuration goes to NM during structural optimization for
Ca$_{0.5}$Na$_{0.5}$Fe$_2$As$_2$ .}
\begin{center}
\begin{tabular}{|c|c|c|c|c|c|c||c|} \hline \hline
   & a & b & c & As(z) &  d$_{FeAs}$ &  M$_{Fe}$ &E (meV) \\ \hline
  \multicolumn{8}{ |c|}{CaFe$2$As$_2$}\\  \hline
NM & 5.63 & 5.63  & {\bf 10.39} &  0.36251  & 2.309  & 0& 0.0\\
AF1 & 5.65  & 5.65 & {\bf 10.60} & 0.36440  & 2.338  & 1.3 & -16\\
AF2 & 5.61 & 5.48  & 11.61 &       0.36695  & 2.367  & 2.2  & -100\\
Exp. & 5.68  & 5.68 & {\bf 11.73} &0.3665 & 2.370  & 1.0 & --\\ \hline
 \multicolumn{8}{| c|}{Ca$_{0.5}$Na$_{0.5}$Fe$_2$As$_2$}\\ \hline 
NM & 5.59  & 5.59 & 10.52 &  0.36284 &   2.31  & 0  & 0.0\\
AF1$^*$ & 5.59 & 5.59  & 10.52 &  0.36284 &  2.31  & 0 & 0.0\\
AF2 & 5.43  & 5.53 & 12.05 &  0.36536  & 2.382  & 2.4  & -97  \\
Exp. & 5.42 & 5.42  & 11.86 &  -- &  --      & 0.0   &  -- \\ \hline
  \multicolumn{8}{ |c|}{BaFe$2$As$_2$}\\  \hline
NM & 5.58 & 5.58  &  12.45 &  0.3479 &    2.319    & 0 & 0.0\\
AF1 & 5.64  & 5.64 & 12.73 &  0.35231 &   2.382    & 2.1 &   -80 \\
AF2 & 5.70 & 5.59  & 12.83 &  0.3549    & 2.408  & 2.4 & -169 \\
Exp.& 5.52 & 5.52 &  13.02 &  0.3545  &   2.397  & 1.0 & -- \\ \hline
  \multicolumn{8}{ |c|}{LaOFeAs}\\  \hline
NM & 5.64 & 5.64  & 8.59  &   0.35944     &  2.332  & 0       & 0.0 \\
AF1 & 5.69  & 5.69  & 8.71 &  0.35128  &     2.393  & 2.1   & -86\\
AF2 & 5.67  &  5.73 & 8.72&   0.34860    &   2.409  & 2.4   & -190 \\
Exp. & 5.70  & 5.70 & 8.737&  0.3479  &      2.407  & 0.35 &  --\\ \hline
\end{tabular}
\end{center}
\end{table}

\begin{figure}
\includegraphics[width=7.5cm]{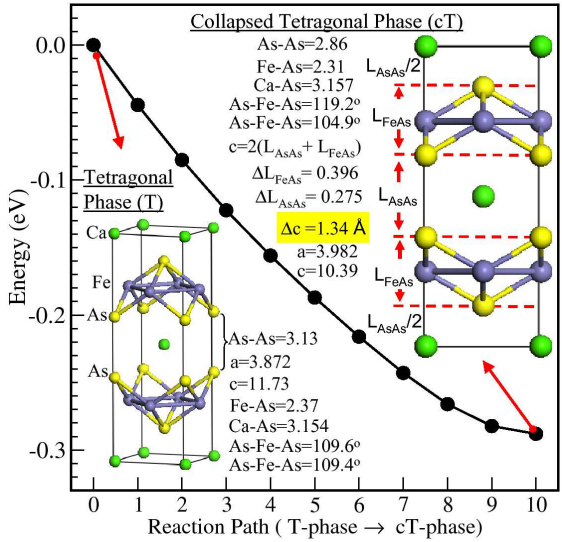} 
\caption{
(color online)
Total energy along a reaction path, showing that the Ca122 tetragonal phase goes  to
collapsed-tetragonal phase without any energy barrier during non-spin polarized
structural optimization. The energy gain  due to 1.34 \AA~
c-axis collapse is about 0.3 eV. The insets show the initial (T-Phase) and final (cT-phase) 
Ca122 structures with relevant bond-distances (in \AA) and angles (in degrees). Note that
in the cT-phase the  change in the height of the FeAs  ($\Delta L_{FeAs}$)
and the As-Ca-As ($\Delta L_{AsAs}$)planes are comparable,
indicating a uniform compression of the whole lattice.
}
\label{fig13}
\end{figure}

Fig.~13 shows an energy reaction path without any energy barrier for 
the collapse of CaFe$_2$As$_2$ with the lost of Fe-spin.
It is surprising that the c-lattice parameter reduces from 11.7 \AA~ to 10.4 \AA~  
but yet the total energy
change is only 0.3 eV per cell (i.e. four Fe ions). One  wonders how
the atoms are rearranged during the collapse of c-axis.  Do the FeAs planes remain 
rigid and get close to each other? Does
this collapse have anything  to do with the effective radius of Fe-ion for different Fe-spin states? Our
answer to these questions is simply no. First of all, the collapse of the c-axis happens rather
uniformly. There is significant and comparable 
decrease in the height of the Fe-As
and As-Ca-As planes, indicating that the whole lattice
almost uniformly shrinks. In our earlier study\cite{yildirim_prl2} 
we traced down the origin of this huge c-axis collapse to 
large As-As interaction between adjacent FeAs planes.

\begin{figure}
\includegraphics[width=6cm]{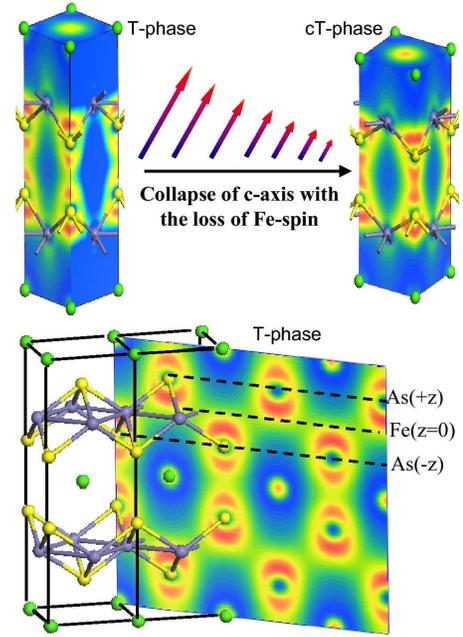}
\caption{
(color online)
Contour plot of one of the orbitals which is responsible for the discovered As-As covalent bonding for
T-phase (top-left) and cT-phase (top-right), respectively. 
Note that the As-As bonding present in both phases is 
much more significant in the cT-phase. 
The bottom panel shows an other orbital on a slice along (110) plane,
indicating clear hybridization between intra-As atoms 
below and above the Fe-plane in the T-phase.   
 }
\label{fig14}
\end{figure}

In order to demonstrate that there are large hybridization between As ions in the Ca122 system, we show
the contour plots of the relevant orbitals in Fig.~14. Again the contour plots in the T- and cT-phases
are quite similar. It is very clear that the As ion below the top Fe-plane makes a bond (or
hybridizes) with the arsenic ion which is above the lower Fe-plane. Hence this overlap of the As-As
along the c-axis makes this system quite isotropic and far from being layered system. From Fig.~14 it is
clear that the As-As bonding along the c-axis got significantly stronger. According to bond-population
analysis, the bond  strength increased almost twice.  Due to close proximity of the As ions in adjacent
Fe-layers, the observation of the As-As interaction is probably not that surprising. What is surprising is  
to see that there is almost the same type of hybridization between two arsenic ions on the same Fe-plane
as shown in the bottom panel of Fig.~14.

Since we have shown that the As ion
above the Fe-plane has a strong overlap with the As ion below the same iron plane, their interaction is
automatically increased as the Fe-As interaction decreases due to decrease in the Fe-moment which
changes the chemistry of the Fe ion. Therefore, we have now a mechanism which explains why the As ion z-values
get shorter with the decreasing Fe-moment. Our mechanism also explains why we see a smaller reduction 
in the c-axis for the LaOFeAs  than the 122 system as listed in Table~1. 
The reduction in the c-axis  in the LaOFeAs system is due to the intra-plane As-As
interaction only since there are no two adjacent FeAs planes to interact with each
other as in the case of Ca122. Our theory also predicts that for larger ions like Ba, we should see less
c-reduction because the As-As distance between two adjacent planes are now larger due to larger ionic
radius of Ba. In Table~1, we also show that similar c-reduction with Fe-spin occurs in the doped
Na$_{0.5}$Ca$_{0.5}$Fe$_2$As$_2$ system  as well.

Since our results suggest that Fe-magnetism is totally lost in the cT-phase, 
one wonders if the $\sim12$~K superconductivity observed in some experiments in the 
vicinity of the collapse cT-phase of  CaFe$_2$As$_2$\cite{pres_cafe2as2}
can be explained by conventional electron-phonon (e-ph) coupling? 
In order to address this question, we have calculated 
phonon spectrum and Eliashberg function from linear response theory\cite{pwscf}. 
We used basically the same method and the equivalent parameters 
that are used in Ref.\onlinecite{ep}  for LaOFeAs. 
Our results are summarized in Fig.~15 and very 
similar to those for LaOFeAs. We obtained a value of 
electron-phonon coupling $\lambda = 0.23$ and the  logarithmically average frequency 
$\omega_{log}= 218$~K, which gives $T_c=0.6$~K using the Allen-Dynes formula with
$\mu^{*}=0$ (i.e. an upper bound for T$_c$).
Hence, if the $\sim12$~K superconductivity observed in some experiments can be confirmed 
to be actual bulk superconductivity in the cT-phase, it would mean that the mechanism of superconductivity 
in the cT-phase of CaFe$_2$As$_2$ is likely unconventional and it has nothing to do with Fe-magnetism which 
is not present in the cT-phase. On the other hand, if the future experiments totally rule out that the cT-phase
of Fe$_2$As$_2$ does not show bulk superconductivity then it would mean that the Fe-magnetism is required
for the superconductivity in these systems. Because of these reasons, the cT and T-phases of CaFe$_2$As$_2$
provide us an invaluable opportunity to study the same system with and without Fe-magnetism. 
We hope that there will
be more focus on the superconducting properties of CaFe$_2$As$_2$ system 
under pressure to resolve the outstanding issues
about the $\sim12$~K superconductivity observed in some experiments.

\begin{figure}
\includegraphics[width=7cm]{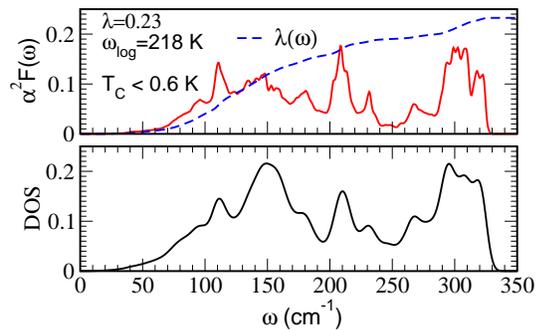}
\caption{
(color online)
Phonon density of states (DOS), Eliashberg function ($\alpha^2F(\omega)$) and the
frequency-dependent e-ph coupling $\lambda(\omega)$ (dashed line) for CaFe$_2$As$_2$ 
in the cT-phase. 
 }
\label{fig15}
\end{figure}

In conclusion, we have revealed surprisingly strong As-As interactions for both intra- and
inter-plane arsenic ions. The strength of this interaction is controlled by the Fe-As chemical bonding. 
Reducing the Fe-moment, reduces the Fe-As bonding, which in turn increases the As-As interaction along
the z-axis, causing arsenic atoms  on opposite sides of Fe-square lattice to move towards   each other. 
This  explains the
high sensitivity of the z-atom positions and the large reduction of the c-axis with 
the Fe-magnetic moment.
From visualization of the electronic orbitals, we show that the 122  systems should not be considered
as layered systems since the As-As inter-plane interaction is as strong as the intra-plane As-As
interaction. The 122 system are in fact quite 3D isotropic than one initially thinks. 
 Since at ambient pressure, we do
not observe large c-axis drops in the superconducting samples, we
conclude that the Fe-magnetic moment
should be present at all times in these systems, 
at least in 122  materials such as CaFe$_2$As$_2$. 
In other words, the iron-pnictide system should be considered as paramagnetic 
(i.e. Fe-moment is present without
long range order). Non-magnetic treatment of Fe ions changes the
 chemistry significantly and is not suitable 
for
description of these systems at ambient pressure. The  giant coupling of the on-site 
Fe-magnetic moment with the As-As bonding that we have discovered here may provide 
a mechanism for the superconductivity.  Since
earlier electron-phonon (ep) coupling 
calculations\cite{ep} were done ignoring the Fe-moment, our results raise some 
questions about the validity of these calculations. 
Currently we are extending such e-ph coupling calculations to
magnetic systems using finite-displacement method in
 which the magnetic response of the system to 
the atomic motion is treated fully unlike the 
standard linear-response perturbation theory.

\section{Magnetic Phonons}

Whenever a new superconductor is discovered, the first thing to do is to check whether
the conventional electron-phonon (e-ph) coupling can explain the observed transition temperature or not. 
This was also the case for Fe-pnictide superconductors. Early electron-phonon 
coupling calculations\cite{ep}
based on standard non-spin polarized perturbation theory indicate that conventional
e-ph can not explain the observed high temperature in Fe-pnictide superconductors. 
The phonon spectrum and its temperature dependence of various 1111 and 122 systems have been also
extensively studied by inelastic neutron scattering\cite{ph_ns_laofeas,ph_ns_laofeas_ty,zbiri},  
inelastic xray scattering\cite{ph_ixs_ndofeas,fukuda,reznik} and
by nuclear-resonance spectrum which is Fe-specific\cite{ph_fedos}. 
These studies did not find any significant
changes in the observed phonon spectrum with superconductivity. However some of these measurements
indicate features which are not produced in the standard linear-response non-magnetic 
phonon calculations. For example, Fukuda {\it at al.}\cite{fukuda} 
found that the calculated phonon DOS agrees 
with the experimental spectrum provided that the computed FeAs force constant is reduced by 30\%.
  Similarly, Reznik {\it et al}\cite{reznik} recently observed that in BaFe$_2$As$_2$ system, 
  the A$_g$ As mode
is around 20-22 meV while there is no such feature in the calculated spectrum. Similar observations
have been made by inelastic neutron scattering\cite{zbiri} where experimetnal DOS has a nice sharp 
peak around 20 meV  while in the calculated spectrum there is nothing in that energy range.  
In addition to these observations, there is also a recent Fe-isotope measurements where an 
isotope coefficient of 0.4 is observed\cite{isotope}. Some anomalous electron-phonon interaction in doped LaOFeAs 
has been also reported from first-principles calculations\cite{ep2}.
All these studies suggest that it is probably too early to 
rule out a possible mechanism based on phonon-mediated superconductivity in Fe-pnictide systems.

\begin{table*}
\caption{The symmetries and energies (in meV) of the optical phonons of LaOFeAS in P4/nmm and Cmma phases.
The  energies of the IR-active modes are  taken from Ref.\onlinecite{dong}. 
The $^{*}$ indicates a significant disagreement. The details of calculations and the 
animations of the modes can be found at 
http://www.ncnr.nist.gov/staff/taner/laofeas}
\begin{center}
\begin{tabular}{|ccc|ccc|ccc|ccc|ccc|} \hline \hline
\multicolumn{15}{|c|}{
$\Gamma (P4/nmm )$ = 2 A$_{1g}$ (R) + 4 A$_{2u}$(IR) + 4 E$_{u}$ (IR) + 4 E$_{g}$(R)  + 2 B$_{1g}$ (R) } \\ 
\multicolumn{15}{|c|}{
$\Gamma (Cmma)$ = 2 A$_{g}$ (R) + 2 B$_{1g}$ (R) +  4 B$_{1u}$(IR) + 4 B$_{2g}$(R) + 4 B$_{2u}$ (IR) + 4 B$_{3g}$(R) + 4 B$_{3u}$ (IR) }\\ \hline
P4/nmm & Cmma & IR  & P4/nmm & Cmma & IR &  P4/nmm & Cmma & IR &  P4/nmm & Cmma & IR & P4/nmm & Cmma & IR \\ \hline
E$_{u}$ 7.3 & 7.4-7.5&  -- & 
A$_{2u}$ 9.9&  10.1& 12.1 & 
E$_{g}$ 14.0& 14.1-14.2& -- & 
E$_{g}$ 17.6& 17.7-17.8& -- & 
A$_{1g}$ 22.1& 22.3& -- \\
A$_{1g}$ 24.9 & 25.1& -- & 
B$_{1g}$ 26.6 & 26.9& -- &  
A$_{2u}$ 31.2 & 31.6& 30.9 & 
E$_{u}$ 33.7 & 34.0-34.1 & 33.2 & 
B$_{1g}$ 35.2& 35.6& -- \\
E$_{g}$ 35.6& 35.9- 36.1& -- & 
 E$_{u}$ {\bf 34.3}& {\bf 34.6- 34.7}& {\bf 42.0$^{*}$ } & 
A$_{2u}$ 48.6 & 49.1& 53.8 &  
E$_{g}$ 51.6& 51.8-52.6& -- &  & & \\ \hline \hline
\end{tabular}
\end{center}
\end{table*}

So far we have shown that the spin-polarized calculations recover from the failure of  non-magnetic
calculations in terms of lattice parameters and internal atomic coordinates. 
 Here we show that magnetic calculations also
resolve the most of the outstanding issues with the observed phonon modes discussed above. 
Our phonon calculations
are done using the plane wave code pwscf within finite displacement technique as described in
Ref.\onlinecite{taner_phonon}. The advantage of direct finite displacement technique over the standard
linear response theory is twofold. The first advantage is that we can do phonon calculations with different
atomic displacements (usually, 0.01, 0.02, and 0.04 \AA) and determine if there is any anharmonic
phonons. For harmonic phonons, the results do not depend on the magnitude of the displacement. 
 The second and probably the most important advantage   is that we treat
 the magnetism and phonon displacements equally and self consistently. We have already seen that
 Fe-magnetic moment is very sensitive to the As-z position and therefore it is not a good
 approximation to assume that the spins are fixed as the atoms move according to a given  phonon
 mode. In our approach, this direct and strong interplay of magnetism and structure is treated self
 consistently.

\begin{figure}
\includegraphics[width=8.0cm]{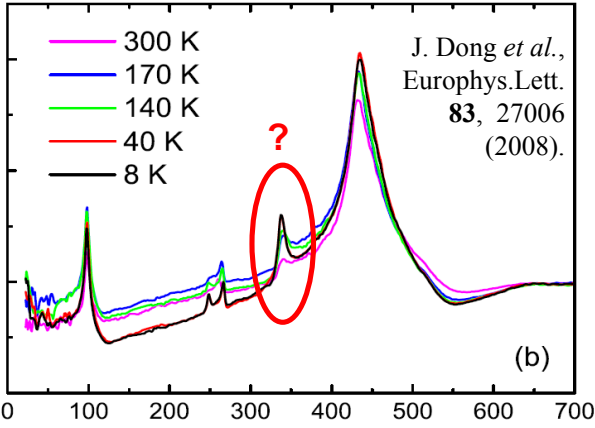}
\caption{
(color online)
The IR data from LaOFEAs sample at different temperatures (adapted from Ref.\cite{dong}), indicating
very strong temperature dependence for the mode near 340 cm$^{-1}$, which is calculated . 
Our calculations do not produce any
zone-center phonon near this energy (see Table.~2). 
}
\label{fig16}
\end{figure}

We first discuss the phonon spectrum of 1111 system (i.e. LaOFeAs) in the high (i.e. paramagnetic)
 and low temperature phases and make some
 comparison with available IR-data which is shown in Fig.~16.  
 Since in the low-T phase, we have both magnetic ordering and
 tetragonal-orthorhombic structural phase transition, it is instructive to study the effect of
 magnetic ordering and structural distortion separately. Hence, we first ignore the Fe-magnetism and
 consider phonon spectrum when the system is non-magnetic with tetragonal and orthorhombic lattice
 parameters. Our results are summarized in Table~2.
From the the symmetry decomposition of the optical phonons in both P4/nmm and
Cmma phases, we note that the distortion does not introduce any new
IR active modes but rather just splits the doubly-degenerate modes into non-degenerate ones.
However the splitting is quite small; the largest is around 0.2 meV. This explains
 why no new modes appear in the
optical measurements after the transition. We also note that the agreement for the 
energies of the zone center phonons with IR data is not as good as one expects.
 In particular, the $E_{u}$ mode observed at 
42 meV is calculated to be 35 meV, a significantly lower value.
 Interestingly, this particular mode has a strong
temperature dependence\cite{dong} as shown in Fig.~16.  
This raises some red flags about the possibility of anharmonic
phonons in these systems. 

\begin{figure}
\includegraphics[width=8.0cm]{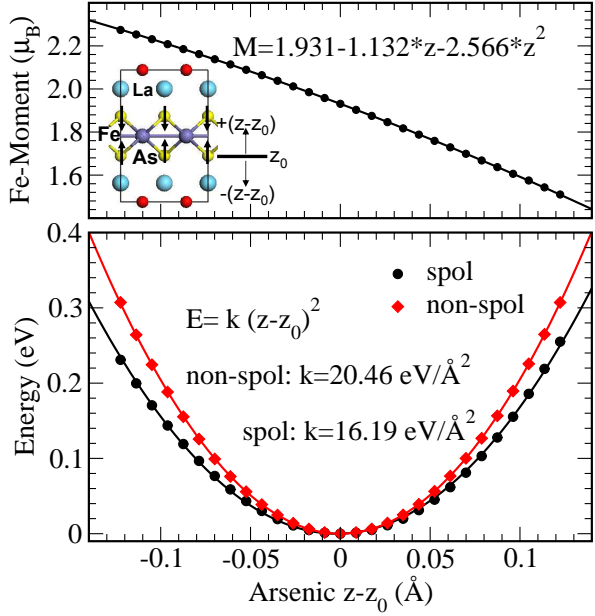}
\caption{
(color online)
Fe-moment and total Energy as As ions are translated along the 
c-axis as shown in the inset. Note that +(z-z0) corresponds to As 
ions moving towards the Fe-plane (z0=0.36). 
Bottom panel indicates that the 
frozen-phonon potential is harmonic and softens about 20\% with 
spin-polarization.
}
\label{fig17}
\end{figure}

Since the As-z position and the Fe-magnetic moment are tightly coupled, one may 
naively expect that the c-polarized
As A$_g$ mode could be quite anharmonic if we perform 
magnetic calculations.  We have checked this first by performing a frozen-phonon type
calculation where we calculate the total energy as the As-atoms are displaced along c-axis in LaOFeAs
system (see Fig.~17). Please note that this is not true A$_g$ mode where both As and La atoms are allowed to move
along c-axis. We will discuss the full phonon calculations next. Figure~17 shows that the total
energy versus the As-displacement can be fit to quadratic equation quite well, 
suggesting this phonon is  harmonic. What is interesting is that the non-spin polarized and spin-polarized
calculations give very different harmonic force constants. 
Including Fe-spin in the calculations softens the
force constant by about 20\%, consistent with  Fukuda's observation\cite{fukuda}
 that one has to soften the
FeAs-force constant to get better agreement with the data. 
However as we shall see below, the renormalization
of the force constants with Fe-spin polarization is quite complicated; almost all of the force
constants  involving Fe and As are basically renormalized.

\begin{figure}
\includegraphics[width=8.0cm]{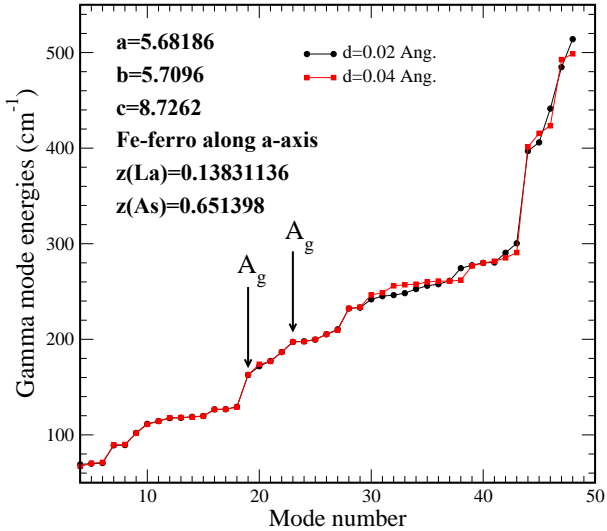}
\caption{
(color online)
Phonon mode energies versus mode numbers  from finite-displacement phonon calculations with two
different displacements, indicating that most of the phonons are harmonic. There are some modes
with very  small
anharmonicity around 250 cm$^{-1}$ which correspond to Fe-Fe stretching modes. The As c-polarized modes with
A$_g$ symmetry are  indicated. These modes are  harmonic, consisting with frozen-phonon energy plot
shown in Fig.~17. 
}
\label{fig18}
\end{figure}

We next check if there are other  modes in LaOFeAs system that could be anharmonic. 
We have carried out full  spin polarized phonon calculations 
(gamma point of the $\sqrt{2}\times\sqrt{2}$ cell) with
atomic displacements of 0.02 \AA \; and 0.04 \AA, respectively. The mode energies are plotted in 
Fig.~18  for both displacements. It is clear that two  As A$_g$ modes are harmonic. There is
some anharmonicity  around 250-300 cm$^{-1}$  corresponding to in-plane Fe-Fe stretching modes.
However this type of mode energy dependence on the atomic displacement
 is typical and does not indicate unusual anharmonicity. Hence, we conclude that in Fe-Pnictide
 system the phonon spectrum is  harmonic. This is quite opposite to what we have in 
 MgB$_2$ superconductor\cite{taner_mgb2} where the in-plane E$_{2g}$ B-B stretching 
 mode was found very anharmonic and responsible for the most  of the e-ph coupling\cite{taner_mgb2}.

We now discuss the effect of the Fe-spin state on phonon modes in LaOFeAs. Our results are also very
similar to 122 systems (in fact the effect of the Fe-spin is more pronounced in 122 systems than in
1111 systems due to strong inter-plane As-As interactions as discussed above). We have
carried out three different phonon calculations. In the first two, we ignore  Fe-magnetism
and just consider  tetragonal (i.e. P4/nmm) and primitive cell of the orthorhombic space group Cmma (i.e.  P2/c). 
The third calculation is fully spin-polarized with the orthorhombic cell parameters. 
We note that when 
iron spins are considered, the space group of the SDW ordered system is no longer Cmma as
reported in neutron experiments\cite{jeff_review}. Cmma is the space group
of the non-magnetic orthorhombic cell.  In our phonon calculations this is a very important point since we use
only symmetry independent displacements to construct the dynamical matrix. 
Using wrong space group, such as Cmma, could average out the
anisotropic force constants due to SDW ordering and give wrong results. 
We determine that the true space group of LaOFeAs when  iron spin is 
considered is Pbmb (spg. number=49, origin 1, a-cb). We note that even
this space group is an approximate since we assume that FeAs planes are ordered ferromagnetically
along c-axis while they are actually ordered antiferromagnetically. However we checked
that the energy difference between ferro and antiferro ordering of the FeAs-planes is too small to
have any significant effect on the calculated phonon spectrum (see $J_\perp$ in Fig.~12 for La111). 
Our results from these three
calculations are  shown in Fig.~19 and Fig.~20. 

\begin{figure}
\includegraphics[width=8.0cm]{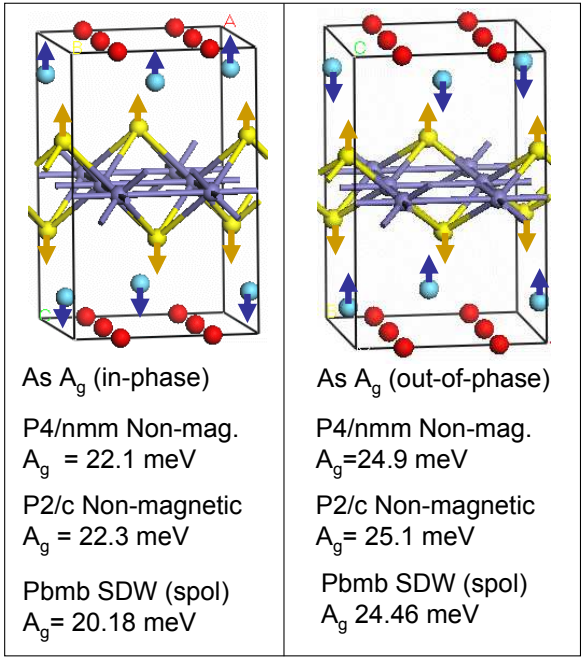}
\caption{
(color online)
Top panel shows two Arsenic c-polarized A$_g$ modes, which are in-phase and out-of-phase with respect to As
and La motions along c-axis. The bottom panel shows the mode energies for non-magnetic tetragonal (P4/nmm),
non-magnetic orthorhombic distorted lattices (P2/c), and 
SDW magnetic configuration (Pbmb).
}
\label{fig19}
\end{figure}

Fig.~19 shows how the Arsenic c-polarized  A$_g$ mode is effected by the structural and magnetic
ordering. We note that due the LaO-plane in 1111 system, we have two A$_g$ modes. In the first one
As- and La-atoms move along c-axis in phase. In the second one As and La atoms move
opposite (i.e. out-of-phase) along the c-axis. Hence in the out-of-phase A$_g$ mode, the
 As-La distance changes as the atoms move. This causes the
out-of-phase A$_g$ mode to have slightly higher energy than in-phase mode. We also note that
 As atoms move twice more than La atoms in the in-phase mode. This is reversed
in the out-of-phase mode. Because of this difference,   the
effect of the Fe-spin state is small on the out-of-phase mode. 
The in-phase A$_g$ mode energy is soften by about 10\% from 22.3
meV to 20.18 meV with the iron spin. This is significant and indicates large magneto-elastic
interactions in these systems as we have already seen in the case of CaFe$_2$As$_2$.

The effect of the spin-polarization on the whole phonon spectrum is shown Fig.~20.
 The phonon density
of states are obtained by calculating the dynamical matrix in a
$2\sqrt{2}\times2\sqrt{2}$ supercell with and without spin-polarization. The  effect of the Fe-spin
on the phonon DOS is significant. As we have already seen, the first effect is softening the
in-phase  As A$_g$ c-polarized mode from 22 meV to 20 meV. The 2nd largest effect occurs near the
36 meV energy range. In the non-spin polarized calculations, we obtain very strong and sharp feature
near 36 meV, inconsistent with the experimental data. This sharp feature near 35-36 meV is 
combination of several modes which are both c-polarized and in-plane oxygen and Fe-Fe modes. 
When we have the Fe-spin included in the calculations, we soften those modes that involve Fe-Fe
stretching  by about \%10, bringing the Fe-modes  
down to 32 meV where the experimental features are observed.

\begin{figure}
\includegraphics[width=8.0cm]{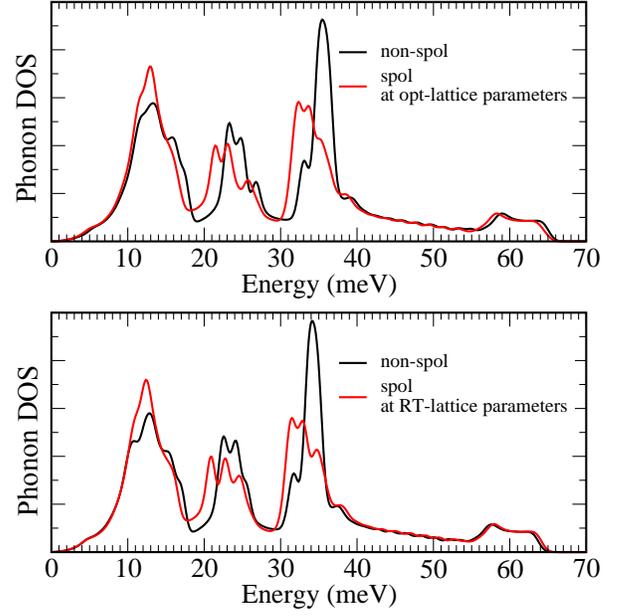}
\caption{
(color online)
 Phonon DOS at optimized orthorhombic cell (top) and room-temperature tetragonal cell 
 (bottom) for non-spin polarized (black) and spin-polarized (red) cases. 
 Note that for non-spol case, we have an intense peak at around 34-35 meV.
  This intense peak is due to oxygen c-polarized  phonons (35.50 meV) and oxygen in-plane phonons (34.76
 meV) as well as Fe-Fe in-plane stretching mode (34.88 meV) and a mixed As c-polarized mode. When
 spin-polarization is included all modes associated with Fe-Fe stretching and As c-modes soften and
 go down to lower energies. 
}
\label{fig20}
\end{figure}

We also checked the real-space force constants obtained from both spin-polarized and non-magnetic
calculations. While La and O onsite force constants are not effected with Fe-spin, the onsite Fe and
As force constants are renormalized by about 10-20\%. For example, the non-spin polarized case gives 
force constants (in eV/$\AA^2$)  (11.06,11.06,8.7) for Fe and (10.48, 10.50, 9.33) for As,
respectively. The spin-polarized calculations renormalize these force constants to (10.8, 8.5, 8.5)
for Fe and (9.2,8.8, 8.6) for As. The change in the force constants are significant. 
 We also observe similar softening up to 20\% in
the Fe-As and Fe-Fe force constants. Hence the 10\% phonon softening of the As $A_g$ mode and Fe-Fe
stretching modes near 35 meV is due to complicated renormalization of the force constants rather than
a simple rescaling of a single force constant as suggested by Futuda {\it et al.}\cite{fukuda}. 
The important point is that Fe-spin is needed to get the observed spectrum even though the
measurements are done on samples at  temperatures well above the T$_N$.

Fig.20 also shows that using room temperature or optimized lattice parameters gives only slightly
different phonon spectrum. This is consistent with the observed weak temperature dependence of the
neutron or inelastic x-ray data. Of course, in reality there is a huge difference between the room
temperature and low temperature calculations if we ignore the Fe-magnetism at room temperature and
consider it at low temperature. 

\begin{figure}
\includegraphics[width=8.0cm]{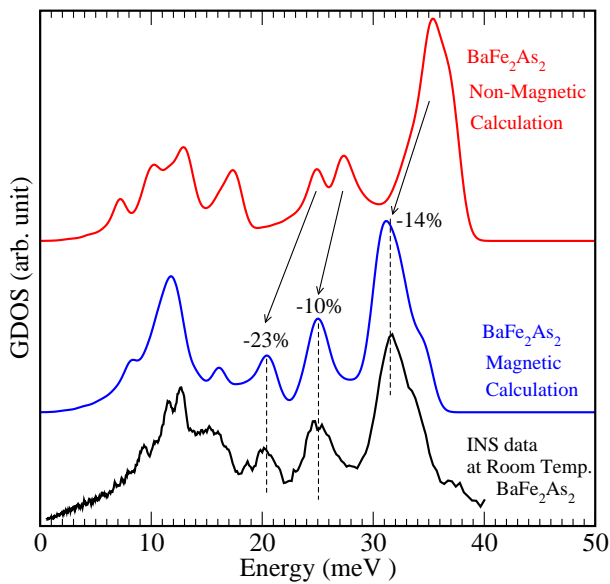}
\caption{
(color online)
Generalized Phonon DOS measured by inelastic neutron scattering
at room temperature for BaFe$_2$As$_2$ (bottom black curve)\cite{zbiri}
 and the
calculated GDOS with (middle blue) and without (top red) Fe-magnetism.
Note that including Fe-spin in the calculations softens the Fe-Fe in-plane
and As c-phonons by about \%10 and \%23 and results a DOS that is in 
perfect agreement with the room temperature data (i.e well above  $T_N = 140$ K). 
}
\label{fig21}
\end{figure}

We will finish this section by presenting  phonon spectrum of Ba$_{0.5}$K$_{0.5}$Fe$_2$As$_2$
 for which  the 
difference between non-magnetic and magnetic DOS is huge as expected from our previous work on 
CaFe$_2$As$_2$ due to strong As-As interactions. In Fig.~21 we compare our calculated phonon DOS with
the neutron data of Zbiri {\it  et al.}\cite{zbiri} on BaFe$_2$As$_2$ sample. 
The authors indicated that a peak near 20 meV can not be produced from non-spin polarized
calculations. 
Zbiri's neutron data shown in Fig.~21 is fully consistent with the
recent inelastic X-ray study of Reznik {\it et al.}\cite{reznik}.
 They have also measure c-polarized As dispersion mode near 
20-22 meV but their non-spin polarized calculations did not produce any peak in the 
energy range of  20-26 meV\cite{reznik}. Zbiri {\it et al.} called this energy range as "pseudo-gap" 
in the phonon spectrum since in the dispersion curves there are no  modes in this energy range
that can give   a sharp peak in the DOS. In Fig.~21 we show that considering magnetic phonons
again solves all the mysterious! Our non-magnetic phonon calculations are in good agreement with 
Reznik phonon dispersion curve\cite{reznik} as well as Zbiri's  results\cite{zbiri}
 and therefore it can not explain the observed GDOS of BaFe$_2$As$_2$. 
 However our magnetic  phonon calculations soften the c-polarized As mode by 
23\%  and the Fe-Fe  modes by about \%10-14, respectively. Hence the magnetic phonon DOS
is now  in perfect agreement with the experimental data. 
This  further supports our theory that Fe-magnetism always 
present in Fe-pnictide systems even at temperatures well above T$_N$.
The iron magnetism could be in the form of fluctuating SDW type 
small magnetic domains\cite{igor_nature} or it could be 
at the atomic limit of paramagnetic Fe ions (i.e. para-magnon).
 More study  is needed to have a better understanding of the Fe-magnetism in
these systems. From the results presented here it is clear that iron-magnetism  in the 
Fe-pnictides is the key factor  that controls  atomic positions, lattice parameters, 
structural phase transition, 
phonon  energies, and most probably the superconducting properties as well. 

\section{conclusions}

Our conclusions can be summarized as follows:

\begin{itemize}
\item
Accurate all-electron fix-spin total energy calculations indicate that 
the ferromagnetic and checkerboard antiferromagnetic 
 ordering in Fe-pnictides were not stable and the stripe Fe-spin configuration 
 (i.e. SDW) is the only stable ground state. Mapping the energies of ferromagnetic, 
 checkerboard AF and SDW spin configurations on to an approximate $J_1^a$-$J_1^b$-$J_2$ model indicates 
 the presence of competing strong antiferromagnetic exchange 
interactions in these systems. This  suggests that  magnetism 
and superconductivity in 
doped Fe-pnictides may be strongly coupled, much like in the high-T$_c$ cuprates.

 \item
  The magnetic stripe SDW phase breaks the tetragonal symmetry,
   removes the frustration, and causes a structural distortion. A simple model based on
   spin-Peierls like transition (i.e. the lattice parameters
   dependence of the exchange constants $J_1$ and $J_2$) is shown to give the correct amount of
   lattice distortion as well as predicts the fine details such as in SDW ordering parallel 
   aligned spin-direction gets shorter while the anti-parallel aligned spin direction gets longer.
   We show that the structural transition temperature, $T_{str}$ does not correlate with FeAs-inter-plane
   magnetic interaction $J_\perp$.  
 The coupling of the magnetic fluctuations to the in-plane strains 
  $\epsilon_{xx}-\epsilon_{yy}$
 and occupations of  orbitals  $d_{xz}-d_{yz}$  may
 change the nature of magnetic transition to first order and splits the magnetic and structural
 ordering temperatures depending on the details of the 
 system\cite{harris_taner_feas,pt_th1,pt_th2,pt_th3,pt_th4}.

 \item
 The magnetic exchange interactions $J_{i,j}(R)$ are calculated as a function of Fe-Fe distance R from
 two totally different approaches. Both methods indicate that the exchange interactions along the
  Fe-Fe square-diagonal spin-direction are short range and antiferromagnetic. Hence it is tempting 
  to conclude that the main diagonal interaction, $J_2$ 
is superexchange type and important contributer towards the stabilization of SDW ordering. 
Along the parallel and antiparallel aligned spin-directions, however, the exchange
 interactions have oscillatory character with an envelop decaying as $1/R^3$, just like in bcc-Fe.
 The major difference between  two methods is that the first nearest-neighbor
 exchange interaction along the parallel spin direction is found to be antiferromagnetic 
 in spin-flip method and
 ferromagnetic  in linear-response theory. Assuming both methods are accurate,
 this implies that   a simple Heisenberg model is
 not appropriate for the Fe-pnictide systems (see Fig.~5).
 However for a given orbital order and spin-configuration, one
 can still use it to describe the low-energy excitations such as spin-waves.

 Since the spin-flip method is more appropriate at  high temperatures above the magnetic phase transition, we obtain
 results that are close to paramagnetic tetragonal symmetry (i.e. $J_1^a \approx J_1^b$) and the system is fully
 frustrated. As the system orders, the occupancy of the $d_{xz}$ start differ from the occupancy of
 $d_{yz}$ orbital, giving rise to orbital-dependent exchange spin-interactions. At low temperatures
 where  spin-wave approximation is
 valid, the exchange interaction along the stripe direction becomes  weak and ferromagnetic.
  At this point, the spin-frustration is totally removed (i.e. all magnetic bonds are satisfied).
  Hence, it would be very interesting to determine the exchange interactions and their temperature 
  dependence by inelastic neutron scattering measurements. We gave a brief discussion about how the spin-wave spectrum 
  would appear from a single crystal sample with orthorhombic twinning.  Due to large  anisotropy of the spin-wave 
  velocities within the ab-plane,  it may be possible to resolve the spin waves along the 
  $a$ and $b$ directions and determine the sign of the $J_1^b$ exchange interaction despite the  
  orthorhombic twinning.

 \item
 We unravel surprisingly strong interactions between arsenic ions in Fe-pnictides,
    the strength of which is controlled by the Fe-spin state in 
    an unprecedented way. Reducing the Fe-magnetic moment, 
    weakens the Fe-As bonding, and in turn, increases As-As interactions, 
    causing a giant reduction in the c-axis. For CaFe$_2$As$_2$ system, 
    this reduction of c-axis with the loss of the Fe-moment is
     as large as 1.4 \AA, an unheard of giant coupling of local
      spin-state of an ion to its lattice. Since the calculated
       large c-reduction has been recently observed only under high-pressure, 
       our results suggest that the iron magnetic moment should be present 
       in Fe-pnictides at all times at ambient pressure. 

\item 
Finally, we showed that Fe-magnetism is also the key in understanding the phonons in Fe-pnictides.
Our magnetic phonon calculations clearly indicate 
that the observed phonon-DOS at room temperature is much
 closer to the calculated magnetic phonon-DOS rather than non-magnetic one.
We find that the in-plane  Fe-Fe and c-polarized As phonon modes are soften  by about 
\%23 and \%10 for 122 and 1111 systems,
respectively, explaining
the observed inealastic x-ray data by Fukuda {\it et al.}\cite{fukuda}
 and by Reznik {\it et al.}\cite{reznik}, and the 
INS room temperature GDOS data  on BaFe$_2$As$_2$ by Zbiri {\it  et al}\cite{zbiri}. 
 This finding further supports our theory that the Fe-magnetism 
 must present in these systems all the time.

\item
The main conclusion of our work is that there is a giant magneto-elastic coupling in Fe-pnictides.
We can successfully predict lattice parameters, atomic positions and phonons in these systems from
first principles  provided that we always
consider Fe-spin in our calculations. Since the current electron-phonon calculations were
carried out without the Fe-spin, it is probably too early to rule out el-phonon coupling as  a
possible mechanism. It is very important that electron-phonon coupling is calculated
self-consistently with the Fe-spin and without the rigid-spin approximation since Fe-moment is very
sensitive to arsenic motion. Currently we are carrying out such calculations and the results will be
presented elsewhere.

\end{itemize}

\section{Acknowledgments}
The author acknowledges fruitful discussions with  P. Dai, A. B. Harris, J. W. Lynn, I. I. Mazin, and W. Ratcliff.

\appendix
\section{Exchange interactions from direct spin-flip method in a large supercell}
We developed a systematic approach where the exchange parameter 
between spin-$i$ and spin-$j$ is
obtained from the total energies of a reference
magnetic configuration and those configurations obtained by  flipping
the spins $i$ and $j$ one at a time and simultaneous 
flipping of both spins. From these four energies, it is possible to
obtain the exchange constant between spin $i$ and $j$.
We note that here we are interested in the isotropic exchange interactions.
We also do not consider spin-orbit interactions
in our calculations. Hence all calculations are done for collinear
spin-configurations.
 
In order to extract superexchange interactions up to a large cutoff
distance, we calculated the total energy for various 
periodic spin configurations based on SDW alignment
of the $z$-components of spin ($S_z = \pm 1$) with a $3\sqrt(2)\times3\sqrt{2}$
supercell of the LaOFeAs which contains 144 atoms (36 of which are Fe ions). Since the spin
configuration is the same from one supercell to the next one we may write
the total energy $E_1$ as

\begin{eqnarray}
E_1 &= & E_0 + \frac{1}{2} \sum_{{\bf R}} \sum_{k,l} 
J(0,k;{\bf R},l) S_k(0)S_l({\bf R}) \nonumber\\
&= & E_0 + \frac{1}{2} \sum_{k,l} K(k,l) S_k S_l \ ,
\end{eqnarray}
where $S(n,{\bf R})\equiv S_n$ is the spin of the $n$th ion in the
supercell at ${\bf R}$ and because of periodicity $K(k,l) = K(l,k) \equiv
\sum_{\bf R} J(k,0;l,{\bf R})$.  It is obvious that we can only
expect to determine $K(k,l)$ and not the individual $J$'s.  However,
since the supercell is reasonably large, we can identify the $K$'s
with the $J$ at the minimal separation. It is also obvious that we can
only hope to determine $K(i,j)$ for $i \neq j$, since the energy
involving $K(i,i)$ depends on $(S_i)^2 = 1$ since $S_i = \pm 1$ for Ni spins.
To determine $K(i,j)$ for
$i \neq j$ we calculate four total energies, $E_1$ and the other three
corresponding energies when we independently change the sign of $S_i$
and $S_j$.  When we change the sign of $S_i$ we get
\begin{eqnarray}
E_2 &= & E_0 + \frac{1}{2} \sum_{k,l} K(k,l) S_k [1 - 2 \delta_{i,k}]
S_l [1- 2 \delta_{i,l}] \ ,
\end{eqnarray}
where $\delta_{n,m}=1$ if $n=m$ and is zero otherwise. Likewise
when we change the sign of $S_j$ we get
\begin{eqnarray}
E_3 &= & E_0 + \frac{1}{2} \sum_{k,l} K(k,l) S_k [1 - 2 \delta_{j,k}]
S_l [1- 2 \delta_{j,l}] \ ,
\end{eqnarray}
and when we change the sign of both spins $i$ and $j$ we get
\begin{eqnarray}
E_4 =  E_0 + \frac{1}{2} \sum_{k,l} && K(k,l) S_k [1 - 2 \delta_{i,k}]
[1 - 2 \delta_{j,k}] S_l  \nonumber \\
&& \times[1 - 2 \delta_{i,l}] [1- 2 \delta_{j,l}] \ .
\end{eqnarray}
Then we construct the quantity $X=E_1 - E_2 - E_3 + E_4$, to get
\begin{eqnarray}
X = \frac{1}{2} \sum_{k,l} K(k,l) S_k S_l &&
[ 2 \delta_{i,k} + 2 \delta_{i,l} - 4 \delta_{i,k} \delta_{i,l}] \times \nonumber \\
&& [ 2 \delta_{j,k} + 2 \delta_{j,l} - 4 \delta_{j,k} \delta_{j,l}] \ .
\end{eqnarray}
Since we require that $i \neq j$, this gives
\begin{eqnarray}
X &=& 4 K(i,j) S_i S_j \ ,
\end{eqnarray}
from which we can extract the value of $K(i,j)$.
If the supercell is large enough,
one can keep only the interactions between the nearest
neighboring supercell images of the spins and therefore
the calculated exchange parameter can be attributed to
spin-interaction between the closest pairs of spins of types
$i$ and $j$.
We also note that there are cases where spin $j$ is at the mid-point
between spin $S_i(0)$ and one of its images at   $S_i({\bf R})$.
In that case, the calculated superexchange constant is twice of the
$J_{i,j}$.  Similarly there are cases where the spin $j$ is at a point
where it interacts equally with four images of the spin $i$. In those cases,
the calculated $J$ is four times $J_{i,j}$.

\end{document}